\documentclass[12pt]{article}
\usepackage{graphicx,epsfig,color,latexsym}
\renewcommand{\theequation}{\arabic{section}.\arabic{equation}}
\newcommand{\bc}{\begin{center}}
\newcommand{\ec}{\end{center}}
\newcommand{\be}{\begin{equation}}
\newcommand{\ee}{\end{equation}}
\newcommand{\bea}{\begin{eqnarray}}
\newcommand{\eea}{\end{eqnarray}}
\newcommand{\ba}{\begin{array}}
\newcommand{\ea}{\end{array}}
\newcommand{\lb}{\label}
\newcommand{\rf}{\ref}
\newcommand{\bfg}{\begin{figure}[htbp]}
\newcommand{\efg}{\end{figure}}
\newcommand{\pr}{Phys. Rev. }
\newcommand{\np}{Nucl. Phys. }

\newcommand{\prp}{Phys. Rep. }
\newcommand{\ap}{Ann. Phys. (N.Y.) }
\newcommand{\pl}{Phys. Lett. }

\newcommand{\rmp}{Rev. Mod. Phys. }
\newcommand{\nc}{Nuovo Cimento }

\begin{document}

\begin{flushright}
IPNO-DR-07-03
\end{flushright}
\vspace{0.5 cm}
\bc
{\large \textbf{Integral equation for gauge invariant \\
\vspace{0.25 cm}
quark two-point Green's function in QCD}}\\
\vspace{1 cm}
H. Sazdjian\\
\textit{IPN, Univ. Paris-Sud 11, CNRS/IN2P3,\\
F-91405 Orsay, France\\
E-mail: sazdjian@ipno.in2p3.fr}
\ec
\par
\renewcommand{\thefootnote}{\fnsymbol{footnote}}
\vspace{0.75 cm}

\bc
{\large Abstract}
\ec
Gauge invariant quark two-point Green's functions defined with
path-ordered gluon field phase factors along skew-polygonal lines
joining the quark to the antiquark are considered. Functional
relations between Green's functions with different numbers of
path segments are established.
An integral equation is obtained for the Green's function defined 
with a phase factor along a single straight line. The equation 
implicates an infinite series of two-point Green's functions, 
having an increasing number of path segments; the related kernels 
involve Wilson loops with contours corresponding to the 
skew-polygonal lines of the accompanying Green's function and with
functional derivatives along the sides of the contours. 
The series can be viewed as an expansion in terms of the global 
number of the functional derivatives of the Wilson loops.   
The lowest-order kernel, which involves a Wilson loop with two 
functional derivatives, provides the framework for an approximate 
resolution of the equation.
\par
\vspace{0.5 cm}
\par

\newpage

\section{Introduction} \lb{s1}
\setcounter{equation}{0}

Gauge invariant objects are expected to provide a more precise 
description of observable quantities than gauge variant ones.
Generally, gauge invariance of multilocal operators is ensured 
with the use of path-ordered phase factors \cite{m1,nm}.
In this respect, the closed loop operator, the so-called Wilson
loop, showed itself a powerful tool for the investigation
of the confinement properties of QCD \cite{w,bw,k}. The properties
of the Wilson loop were studied in detail in a long series of papers
\cite{p,mm1,mm2,mgd,mk,m2,g,kkk,dv,bnsg}. 
\par
On the other hand, the usual machinery of quantum field theory, based 
on the Dyson--Schwin\-ger integral equations \cite{d,schw1}, does not 
apply in a straight\-forward way to Green's functions of operators involving
path-ordered phase factors. The main reason is related to the difficulty
of obtaining the functional inverses of the nonlocal gauge invariant
Green's functions and thus of being able to define analogues of 
proper vertices, which play a crucial role in the formulation of 
integral equations. Expressions of gauge invariant quark-antiquark 
Green's functions in terms of Wilson loops were obtained in the past 
\cite{bmpbbp,bcpbmp,sdks} with the use of the Feynman--Schwinger 
representation of the quark propagator \cite{f,schw2,nw,stj}; these, 
however, could not be transformed into equivalent integral equations 
without the recourse to approximations related to the quark motion.
\par
The purpose of the present paper is to investigate the possibilities 
of deriving integral or integro-differential equations for gauge
invariant Green's functions which might allow for a systematic study 
of their various properties. We concentrate in this work on the
quark gauge invariant two-point function, in which the quark and the 
antiquark fields are joined by a path-ordered phase factor, but the 
methods which we shall develop are readily applicable to more general 
cases. 
\par
Our starting point is a particular representation of the
quark propagator in the presence of an external gluon field, where
it is expressed as a series of terms involving path-ordered phase
factors along successive straight lines forming generally 
skew-polygonal lines.
That representation is a relativistic generalization of the one 
introduced by Eichten and Feinberg in the nonrelativistic case
\cite{ef}; it was already used in a previous work for deriving a
bound state equation for quark-antiquark systems \cite{js}; however,
in the latter work, the bound state equation was derived by
circumventing the explicit writing of an integral equation for the
related Green's function and of the neglected higher-order terms of 
the interaction kernel. One of the main properties of the above 
representation is that in gauge invariant quantities, at each order 
of the expansion, the paths of the phase factors close up to form 
a Wilson loop. Thus, the corresponding Green's function becomes 
expressed, through a series expansion, in terms, among others, of 
Wilson loops having skew-polygonal contours with an increasing number 
of sides.
\par  
Several differences occur with respect to the formulation of the
Dyson--Schwinger equations. First, for the reasons mentioned above, 
proper vertices are not introduced; instead, we work directly with
Green's functions; the various kernels that appear are written 
explicitly in terms of functional derivatives of the logarithm of 
the Wilson loop average and of the quark Green's function. Second, 
starting from the simplest gauge invariant two-point function, 
constructed with a phase factor along a single straight line joining 
the quark to the antiquark, one generates, through the equations of 
motion, a chain of new gauge invariant two-point functions with phase 
factors along $n$-sided skew-polygonal lines between the quark and the 
antiquark ($n>1$). On the other hand, every such Green's function 
(with $n$ skew-polygonal sides) can be related with the aid of 
functional relations to the lowest-order Green's function ($n=1$) and 
thus, in principle, an equation involving only the latter Green's 
function is possible to construct. The third difference arises at the 
level of the presence of nested kernels, which do not occur in the 
Dyson--Schwinger equations and which persist here due to background 
effects induced by the Wilson loops: each nested kernel is modified 
by its new background when inserted inside a higher-order Wilson loop 
as compared to its original expression. The remaining terms in the 
kernels have the property of conventional irreducibility.
\par
The integral equation that we obtain is constructed as
an expansion in terms of the global number of derivatives of the
logarithm of the Wilson loop average. Although it involves an 
infinite series of kernels and Green's functions, at each
order of the expansion the explicit expressions of the kernels and
of the relations between high-order Green's functions with the 
lowest-order one can be obtained from definite formulas.
\par 
On practical grounds, an increasing number of derivatives of a Wilson
loop, each derivative occurring on a different region of the contour,
is generally expected to give a relatively decreasing contribution at
short- and at large-distances. Therefore, the series expansion of the 
kernels in terms of functional derivatives of Wilson loops can also be
considered as a perturbative expansion, the most important contribution 
coming from the lowest-order non-vanishing term and involving the 
smallest number of derivatives. That property allows us to consider
solving the integral equation with appropriate approximations.
\par
The plan of the paper is the following. In Sec. 2, we introduce the
definitions and conventions that will be used throughout this work.
Section 3 deals with the representation of the quark propagator in
external field in terms of path-ordered phase factors. In Sec. 4, 
functional relations are established between various Green's functions.
In Sec. 5, the integral equation for the gauge invariant quark 
two-point function with a straight line path is established and 
the structure of the kernel terms is displayed. Section 6 deals 
with the question of analyticity properties of the Green's function. 
A summary and comments follow in Sec 7. Two appendices are devoted to 
the presentation of the summation method with free propagators and 
the study of the self-energy function.
\par

\section{Definitions and conventions} \lb{s2}
\setcounter{equation}{0}

We introduce in this section the main definitions and conventions
that we shall use throughout this work.
\par 
We consider a path-ordered phase factor along a line $C_{yx}$
joining a point $x$ to a point $y$, with an orientation defined
from $x$ to $y$:
\be \lb{2e1}
U(C_{yx};y,x)\equiv U(y,x)=Pe^{{\displaystyle -ig\int_x^y 
dz^{\mu}A_{\mu}(z)}},
\ee
where $A_{\mu}=\sum_a A_{\mu}^at^a$, $A_{\mu}^a$ ($a=1,\ldots,N_c^2-1$) being
the gluon fields and $t^a$ the generators of the gauge group $SU(N_c)$ in the
fundamental representation, with the normalization
tr$t^at^b=\frac{1}{2}\delta^{ab}$. A more detailed  definition of $U$
is given by the series expansion in the coupling constant $g$; all
equations involving $U$ can be obtained from the latter expression. 
Parametrizing the line $C$ with a parameter $\lambda$, $C=\{x(\lambda)\}$,
$0\leq \lambda \leq 1$, such that $x(0)=x$ and $x(1)=y$, a variation of $C$
induces the following variation of
$U$ [$U(x(\lambda),x(\lambda'))\equiv U(\lambda,\lambda')$,
$A(x(\lambda))\equiv A(\lambda)$]:
\bea \lb{2e2}
\delta U(1,0)&=&-ig\delta x^{\alpha}(1)A_{\alpha}(1)U(1,0)
+igU(1,0)A_{\alpha}(0)\delta x^{\alpha}(0)\nonumber \\
& &+ig\int_0^1d\lambda U(1,\lambda)x^{\prime \beta}(\lambda)F_{\beta\alpha}
(\lambda)\delta x^{\alpha}(\lambda)U(\lambda,0),
\eea
where $x'=\frac{\partial x}{\partial \lambda}$ and $F$ is the field
strength, $F_{\mu\nu}=\partial_{\mu}A_{\nu}-\partial_{\nu}A_{\mu}
+ig[A_{\mu},A_{\nu}]$. The variations inside the integral lead to
functional differentiation of $U$, while the variations at the end points 
(marked points) are defined as leading to ordinary differentiation. 
The functional derivative of $U$ with respect to $x(\lambda)$ 
($0< \lambda <1$) is then \cite{p}:
\be \lb{2e3}
\frac{\delta U(1,0)}{\delta x^{\alpha}(\lambda)}=
igU(1,\lambda)x^{\prime \beta}(\lambda)F_{\beta\alpha}(\lambda)U(\lambda,0).
\ee
\par
For paths defined along rigid lines, the variations inside the integral
in Eq. (\rf{2e2}) are related, with appropriate weight factors, to those 
of the end points; a displacement of one end point generates a displacement 
of the whole line with the other end point left fixed. Considering now
a rigid straight line between $x$ and $y$, an ordinary derivation at 
the end points yields:
\bea 
\lb{2e4a}
& &\frac{\partial U(y,x)}{\partial y^{\alpha}}=-igA_{\alpha}(y)U(y,x)
+ig(y-x)^{\beta}\int_0^1d\lambda\,\lambda\, U(1,\lambda)F_{\beta\alpha}
(\lambda)U(\lambda,0),\\
\lb{2e4b}
& &\frac{\partial U(y,x)}{\partial x^{\alpha}}=+igU(y,x)A_{\alpha}(x)
+ig(y-x)^{\beta}\int_0^1d\lambda\,(1-\lambda)\, U(1,\lambda)F_{\beta\alpha}
(\lambda)U(\lambda,0).\nonumber \\
& &
\eea
\par
When considering path variations of gauge invariant quantities, with
paths made of segments, the end point contributions involving the 
explicit $A_{\alpha}$ terms disappear, being cancelled by similar 
contributions coming from neighboring segments or from variations 
of neighboring fields. The general contributions that remain at the 
end are those coming from the internal part of the segments
represented by the integrals in Eqs. (\rf{2e4a})-(\rf{2e4b}). We adopt 
the following conventions to represent such contributions:
\bea 
\lb{2e5a}
& &\frac{\bar\delta U(y,x)}{\bar\delta y^{\alpha +}}\equiv
ig(y-x)^{\beta}\int_0^1d\lambda\,\lambda\, U(1,\lambda)F_{\beta\alpha}
(\lambda)U(\lambda,0),\\
\lb{2e5b}
& &\frac{\bar\delta U(y,x)}{\bar\delta x^{\alpha -}}\equiv
ig(y-x)^{\beta}\int_0^1d\lambda\,(1-\lambda)\, U(1,\lambda)F_{\beta\alpha}
(\lambda)U(\lambda,0).
\eea
The first equation above corresponds to a displacement of the end
point of the segment (taking into account the orientation on the path),
while the second equation corresponds to a displacement of the starting
point of the segment. Equations (\rf{2e4a}) and (\rf{2e4b}) can be
written as
\bea
\lb{2e6}
& &\frac{\partial U(y,x)}{\partial y^{\alpha}}=-igA_{\alpha}(y)U(y,x)
+\frac{\bar\delta U(y,x)}{\bar\delta y^{\alpha +}},\\
\lb{2e7}
& &\frac{\partial U(y,x)}{\partial x^{\alpha}}=+igU(y,x)A_{\alpha}(x)
+\frac{\bar\delta U(y,x)}{\bar\delta x^{\alpha -}}.
\eea
\par
If two phase factors $U(z,y)$ and $U(y,x)$ along segments are joined 
at the point $y$ (a marked point), then, with the aid of the previous 
notations, we have:
\be \lb{2e8}
\frac{\partial}{\partial y^{\alpha}}\Big(U(z,y)U(y,x)\Big)=
\frac{\bar\delta U(z,y)}{\bar\delta y^{\alpha -}}U(y,x)
+U(z,y)\frac{\bar\delta U(y,x)}{\bar\delta y^{\alpha +}}.
\ee
\par 
The Wilson loop, denoted $\Phi(C)$, is defined as the trace in color space
of the path-ordered phase factor (\rf{2e1}) along a closed contour $C$:
\be \lb{2e9}
\Phi(C)=\frac{1}{N_c}\mathrm{tr}Pe^{{\displaystyle -ig\oint_C
dx^{\mu}A_{\mu}(x)}},
\ee
where the factor $1/N_c$ has been put for normalization. It is a gauge
invariant quantity. Its vacuum expectation value is denoted $W(C)$:
\be \lb{2e10}
W(C)=\langle\Phi(C)\rangle,
\ee
the averaging being defined in the path integral formalism.
\par 
We shall represent the Wilson loop average as an exponential function,
whose argument is a functional of the contour $C$ \cite{mm1,dv}:
\be \lb{2e12}
W(C)=e^{{\displaystyle F(C)}}.
\ee
In perturbation theory, $F(C)$ is given by the sum of all connected
diagrams, the connection being defined with respect to the contour $C$,
after subtraction of reducible parts \cite{dv}.
Variations of $W(C)$ due to local deformations of the contour $C$ can 
then be expressed in terms of variations of $F(C)$:
\be \lb{2e12a}
\frac{\delta W(C)}{\delta x^{\alpha}}\bigg|_{x\in C}= 
\frac{\delta F(C)}{\delta x^{\alpha}}\bigg|_{x\in C}\,W(C).
\ee
This property is also generalized to the case of rigid variations
of paths (segments). 
If the contour $C$ is a skew-polygon $C_n$ with $n$ sides and $n$ 
successive marked points $x_1$, $x_2$, $\ldots$, $x_n$ at the cusps, 
then we write:
\be \lb{2e13}
W(x_n,x_{n-1},\ldots,x_1)=W_n=
e^{{\displaystyle F_n(x_n,x_{n-1},\ldots,x_1)}}
=e^{{\displaystyle F_n}},
\ee
the orientation of the contour going from $x_1$ to $x_n$ through
$x_2$, $x_3$, etc. (i.e., towards $x$s with indices increasing by
one unit). Then, according to the definitions (\rf{2e5a})-(\rf{2e8}), 
the notation $\bar\delta F_n/\bar\delta x_i^- $ means that the 
derivation acts on the internal part of the segment $x_ix_{i+1}$ 
with $x_{i+1}$ held fixed ($x_{n+1}=x_1$), while 
$\bar\delta F_n/\bar\delta x_i^+$ means that the derivation acts on the 
internal part of the segment $x_{i-1}x_i$ with $x_{i-1}$ held fixed 
($x_0=x_n$).
\par
The gauge invariant two-point quark Green's function is defined as
\be \lb{2e14}
S_{\alpha\beta}(x,x';C_{x'x})=-\frac{1}{N_c}\,\langle\overline 
\psi_{\beta}(x')\,U(C_{x'x};x',x)\,\psi_{\alpha}(x)\rangle,
\ee
$\alpha$ and $\beta$ being the Dirac spinor indices, while the 
color indices are implicitly summed. In the present work we shall
mainly deal with paths along skew-polygonal lines. For such lines 
with $n$ sides and $n-1$ junction points $y_1$, $y_2$, $\ldots$, 
$y_{n-1}$ between the segments, we define:
\be \lb{2e15}
S_{(n)}(x,x';y_{n-1},\dots,y_1)=-\frac{1}{N_c}\,\langle\overline \psi(x')
U(x',y_{n-1})U(y_{n-1},y_{n-2})\ldots U(y_1,x)\psi(x)\rangle.
\ee
The simplest such function corresponds to $n=1$, for which the points
$x$ and $x'$ are joined by a single straight line:
\be \lb{2e16}
S_{(1)}(x,x')\equiv S(x,x')=-\frac{1}{N_c}\,\langle\overline \psi(x')
\,U(x',x)\,\psi(x)\rangle.
\ee
(We shall generally omit the index 1 from that function.)
\par
The free propagator will be designated by $S_0$ (without color group 
content):
\be \lb{2e17}
S_0(x,x')=S_0(x-x')=\int \frac{d^4p}{(2\pi)^4}\,
e^{{\displaystyle -ip.(x-x')}}\,\frac{i}{\gamma.p-m+i\varepsilon}.
\ee
\par
In conjunction with the definitions (\rf{2e5a})-(\rf{2e5b}), we shall
also introduce the notations
\bea \lb{2e18}
& &\frac{\bar\delta S(x,x')}{\bar\delta x^{\mu -}}=
-\frac{1}{N_c}\,\langle\overline \psi(x')\,
\frac{\bar\delta U(x',x)}{\bar\delta x^{\mu -}}\,\psi(x)\rangle,
\ \ \ \frac{\bar\delta S(x,x')}{\bar\delta x^{'\nu +}}=
-\frac{1}{N_c}\,\langle\overline \psi(x')\,
\frac{\bar\delta U(x',x)}{\bar\delta x^{'\nu +}}\,\psi(x)\rangle.
\nonumber \\
& &
\eea
\par

\section{The quark propagator in external field} \lb{s3}
\setcounter{equation}{0}


We shall use a two-step quantization method, by first integrating
the quark fields and then, at a second stage, integrating the
gluon fields through Wilson loops. The first operation yields
among various quantities the quark propagator in 
the presence of an arbitrary external gluon field. The latter,
designated by $S(x,x';A)$, or by $S(A)$ for short, satisfies the usual 
equation
\be \lb{3e1}
\Big(i\gamma.\partial_{(x)}-m-g\gamma.A(x)\Big)S(x,x';A)=i\delta^4(x-x').
\ee
To exhibit a Wilson loop structure in gauge invariant quantities, it
is necessary to describe the quark propagator in external field by 
means of path-ordered phase factors. To this
end, we shall first introduce a representation, already used in Ref. 
\cite{js}, which combines path-ordered phase factors along straight 
lines and free quark propagators. At a later stage, to sum directly 
self-energy effects, we shall replace the free quark propagator by 
the full gauge invariant Green's fuction (\rf{2e16}). 
\par
The starting point of the representation is the gauge covariant 
composite object, denoted $\widetilde S_0(x,x')$, made of a free fermion 
propagator $S_0(x,x')$ (without color group content) multiplied by the 
path-ordered phase factor $U(x,x')$ [Eq. (\rf{2e1})] taken along the 
straight line $xx'$:
\be \lb{3e2}
\Big[\widetilde S_0(x,x')\Big]_{\ b}^a\equiv 
S_0(x,x')\Big[U(x,x')\Big]_{\ b}^a.
\ee
[$a,b$: color indices.]
The advantage of the straight line over other types of line is that 
under Lorentz transformations it remains form invariant and in
the limit $x'\rightarrow x$ $U$ tends to unity in an unambiguous way.
$\widetilde S_0$ satisfies the following equation with respect to $x$:
\be \lb{3e3}
\Big(i\gamma.\partial_{(x)}-m-g\gamma.A(x)\Big)\widetilde S_0(x,x')=
i\delta^4(x-x')+i\gamma^{\alpha}\frac{\bar\delta U(x,x')}
{\bar\delta x^{\alpha +}}S_0(x,x').
\ee
A similar equation also holds with respect to $x'$, with $x$ held fixed,
with the Dirac and color group matrices acting from the right.
\par
The quantity $-i(i\gamma.\partial_{(x)}-m-g\gamma.A(x))\delta^4(x-x')$ is
the inverse of the quark propagator $S(x,x';A)$ in the presence of the 
external gluon field $A$. Reversing Eq. (\rf{3e3}) with respect to 
$S(A)^{-1}$, one obtains an equation for $S(A)$ in terms of 
$\widetilde S_0$:
\be \lb{3e4}
S(x,x';A)=\widetilde S_0(x,x')-\int d^4x''\,S(x,x'';A)\,\gamma^{\alpha}\,
\frac{\bar\delta \widetilde S_0(x'',x')}{\bar\delta x^{''\alpha +}}.
\ee
Using the equation with $x'$, or making in Eq. (\rf{3e4}) an integration 
by parts, one obtains another equivalent equation:
\be \lb{3e5}
S(x,x';A)=\widetilde S_0(x,x')+\int d^4x''\,
\frac{\bar\delta \widetilde S_0(x,x'')}{\bar\delta x^{''\alpha -}}\,
\gamma^{\alpha}\,S(x'',x';A).
\ee
\par
Equations (\rf{3e4}) or (\rf{3e5}) allow us to obtain the propagator
$S(A)$ as an iteration series with respect to $\widetilde S_0$, which
contains the free fermion propagator, by maintaining at each order
of the iteration its gauge covariance property. For instance, the 
expansion of Eq. (\rf{3e4}) takes the form:
\bea \lb{3e6}
S(x,x';A)&=&\widetilde S_0(x,x')-\int d^4y_1\,\widetilde S_0(x,y_1)\,
\gamma^{\alpha_1}\,\frac{\bar\delta \widetilde S_0(y_1,x')}
{\bar\delta y_1^{\alpha_1 +}}\nonumber \\
& &\ \ +
\int d^4y_1d^4y_2\,\widetilde S_0(x,y_1)\,\gamma^{\alpha_1}\,
\frac{\bar\delta \widetilde S_0(y_1,y_2)}
{\bar\delta y_1^{\alpha_1 +}}\,
\gamma^{\alpha_2}\,\frac{\bar\delta \widetilde S_0(y_2,x')}
{\bar\delta y_2^{\alpha_2 +}}+\cdots\ .
\eea
Equations (\rf{3e4}) and (\rf{3e5}) are relativistic generalizations 
of the representation used for heavy quark propagators starting from 
the static case \cite{ef}.
\par
In order to sum, for later purposes, self-energy effects, one can
use for the expansion of the propagator $S(A)$, instead of the free
propagator $S_0$, the full gauge invariant Green's function (\rf{2e16}).
To this end, we define a generalized version of the gauge covariant
object $\widetilde S_0$ [Eq. (\rf{3e2})], by replacing in it $S_0$
with $S$ [Eq. (\rf{2e16})]:
\be \lb{3e7}  
\Big[\widetilde S(x,x')\Big]_{\ b}^a\equiv 
S(x,x')\Big[U(x,x')\Big]_{\ b}^a.
\ee
The Green's function $S$ satisfies the following equations of motion:
\bea 
\lb{3e8a}
& &(i\gamma.\partial_{(x)}-m)S(x,x')=i\delta^4(x-x')
+i\gamma^{\mu}\,\frac{\bar\delta S(x,x')}{\bar\delta x^{\mu -}},\\
\lb{3e8b}
& &S(x,x')(-i\gamma.\stackrel{\leftarrow}{\partial}_{(x')}-m)=
i\delta^4(x-x')
-i\frac{\bar\delta S(x,x')}{\bar\delta x^{'\mu +}}\,\gamma^{\mu}.
\eea
[Notice that the orientation of the path in $S(x,x')$ is from $x$ to $x'$.]
Then $\widetilde S$ satisfies the equation
\bea \lb{3e9}
& &\Big(i\gamma.\partial_{(x)}-m-g\gamma.A(x)\Big)\widetilde S(x,x')=
i\delta^4(x-x')\nonumber \\
& &\ \ \ \ \ +i\gamma^{\alpha}\Big(\frac{\bar\delta S(x,x')}
{\bar\delta x^{\alpha -}}U(x,x')+S(x,x')\frac{\bar\delta U(x,x')}
{\bar\delta x^{\alpha +}}\Big),
\eea
from which one deduces the expansion of $S(A)$ around $S$: 
\bea \lb{3e10}
& &S(x,x';A)=S(x,x')U(x,x')-S(x,y;A)\,\gamma^{\alpha}\,
\Big(\frac{\bar\delta S(y,x')}{\bar\delta y^{\alpha -}}U(y,x')
+S(y,x')\frac{\bar\delta U(y,x')}{\bar\delta y^{\alpha +}}\Big).
\nonumber \\
& &
\eea
[The integrations on intermediate points are implicit.]
Using the equations of $S$ and $S(A)$ relative to $x'$, or making in Eq. 
(\rf{3e10}) an integration by parts, one obtains another equivalent 
equation:
\be \lb{3e11}
S(x,x';A)=S(x,x')U(x,x')+
\Big(\frac{\bar\delta S(x,y)}{\bar\delta y^{\alpha +}}U(x,y)
+S(x,y)\frac{\bar\delta U(x,y)}{\bar\delta y^{\alpha -}}\Big)
\,\gamma^{\alpha}\,S(y,x';A).
\ee
A graphical representation of Eq. (\rf{3e11}) is shown in 
Fig. \rf{3f1}
\par
\bfg
\bc
\begin{picture}(0,0)%
\includegraphics{3f1.pstex}%
\end{picture}%
\setlength{\unitlength}{2960sp}%
\begingroup\makeatletter\ifx\SetFigFont\undefined%
\gdef\SetFigFont#1#2#3#4#5{%
  \reset@font\fontsize{#1}{#2pt}%
  \fontfamily{#3}\fontseries{#4}\fontshape{#5}%
  \selectfont}%
\fi\endgroup%
\begin{picture}(9532,1723)(826,-4703)
\put(3001,-4336){\makebox(0,0)[lb]{\smash{{\SetFigFont{11}{13.2}{\familydefault}{\mddefault}{\updefault}{\color[rgb]{0,0,0}$=$}%
}}}}
\put(7801,-4336){\makebox(0,0)[lb]{\smash{{\SetFigFont{11}{13.2}{\familydefault}{\mddefault}{\updefault}{\color[rgb]{0,0,0}$+$}%
}}}}
\put(1426,-4636){\makebox(0,0)[lb]{\smash{{\SetFigFont{11}{13.2}{\familydefault}{\mddefault}{\updefault}{\color[rgb]{0,0,0}$S(A)$}%
}}}}
\put(4726,-4036){\makebox(0,0)[lb]{\smash{{\SetFigFont{11}{13.2}{\familydefault}{\mddefault}{\updefault}{\color[rgb]{0,0,0}$U$}%
}}}}
\put(5251,-4336){\makebox(0,0)[lb]{\smash{{\SetFigFont{11}{13.2}{\familydefault}{\mddefault}{\updefault}{\color[rgb]{0,0,0}$+$}%
}}}}
\put(4126,-4636){\makebox(0,0)[lb]{\smash{{\SetFigFont{11}{13.2}{\familydefault}{\mddefault}{\updefault}{\color[rgb]{0,0,0}$S$}%
}}}}
\put(8926,-3436){\makebox(0,0)[lb]{\smash{{\SetFigFont{14}{16.8}{\familydefault}{\mddefault}{\updefault}{\color[rgb]{0,0,0}$+$}%
}}}}
\put(6301,-3661){\makebox(0,0)[lb]{\smash{{\SetFigFont{14}{16.8}{\familydefault}{\mddefault}{\updefault}{\color[rgb]{0,0,0}$+$}%
}}}}
\put(826,-4486){\makebox(0,0)[lb]{\smash{{\SetFigFont{9}{10.8}{\familydefault}{\mddefault}{\updefault}{\color[rgb]{0,0,0}$x$}%
}}}}
\put(2551,-4486){\makebox(0,0)[lb]{\smash{{\SetFigFont{9}{10.8}{\familydefault}{\mddefault}{\updefault}{\color[rgb]{0,0,0}$x'$}%
}}}}
\put(3526,-4486){\makebox(0,0)[lb]{\smash{{\SetFigFont{9}{10.8}{\familydefault}{\mddefault}{\updefault}{\color[rgb]{0,0,0}$x$}%
}}}}
\put(4951,-4486){\makebox(0,0)[lb]{\smash{{\SetFigFont{9}{10.8}{\familydefault}{\mddefault}{\updefault}{\color[rgb]{0,0,0}$x'$}%
}}}}
\put(5626,-4486){\makebox(0,0)[lb]{\smash{{\SetFigFont{9}{10.8}{\familydefault}{\mddefault}{\updefault}{\color[rgb]{0,0,0}$x$}%
}}}}
\put(7426,-4486){\makebox(0,0)[lb]{\smash{{\SetFigFont{9}{10.8}{\familydefault}{\mddefault}{\updefault}{\color[rgb]{0,0,0}$x'$}%
}}}}
\put(8326,-4486){\makebox(0,0)[lb]{\smash{{\SetFigFont{9}{10.8}{\familydefault}{\mddefault}{\updefault}{\color[rgb]{0,0,0}$x$}%
}}}}
\put(10051,-4486){\makebox(0,0)[lb]{\smash{{\SetFigFont{9}{10.8}{\familydefault}{\mddefault}{\updefault}{\color[rgb]{0,0,0}$x'$}%
}}}}
\put(9376,-3136){\makebox(0,0)[lb]{\smash{{\SetFigFont{9}{10.8}{\familydefault}{\mddefault}{\updefault}{\color[rgb]{0,0,0}$y$}%
}}}}
\put(6676,-3136){\makebox(0,0)[lb]{\smash{{\SetFigFont{9}{10.8}{\familydefault}{\mddefault}{\updefault}{\color[rgb]{0,0,0}$y$}%
}}}}
\end{picture}%

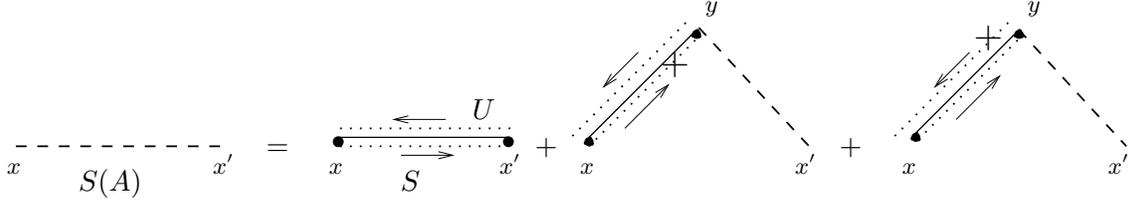
\caption{Graphical representation of Eq. (\rf{3e11}). The double line
(one full and one dotted joining two circles) represents the gauge 
invariant Green's function $S_{(1)}\equiv S$ [Eq. (\rf{2e16})] with a 
path along a single straight line; the single dotted line represents 
the phase factor, the dashed line the quark propagator in the external 
gluon field, the arrow the orientation on the path. The cross represents 
the rigid path derivation with one end fixed; it is placed near the end 
point which is submitted to derivation. $y$ is an integration variable.}
\lb{3f1}
\ec
\efg
Equations (\rf{3e10})-(\rf{3e11}) constitute the basic formulas 
that will be used to express equations of motion of gauge invariant 
quark Green's functions in terms of Wilson loops and gauge invariant
two-point Green's functions.
\par

\section{Functional relations for Green's functions} \lb{s4}
\setcounter{equation}{0}


Functional relations between various gauge invariant quark Green's
functions are ob\-tain\-ed with a systematic use of Eqs. (\rf{3e10}) or 
(\rf{3e11}).
\par
Let us consider the Green's function $S_{(n)}$ [Eq. (\rf{2e15})].
Integrating with respect to the quark fields, one obtains:
\be \lb{4e1}
S_{(n)}(x,x';y_{n-1},\ldots,y_1)=\frac{1}{N_c}\,\langle U(x',y_{n-1})
U(y_{n-1},y_{n-2})\cdots U(y_1,x)S(x,x';A)\rangle.
\ee 
The simplest case of this equation, corresponding to $n=1$, is:
\be \lb{4e2}
S_{(1)}(x,x')\equiv S(x,x')=\frac{1}{N_c}\,\langle U(x',x)S(x,x';A)\rangle.
\ee
The quark field integration yields also a corresponding
determinant, which is a functional of the quark propagator in 
the external gluon field $A$. That determinant will not, however, play 
an active role in the subsequent calculations and hence will not 
explicitly appear in the various formulas that we shall meet; it 
will rather contribute as a background effect; in particular, it 
contributes to the evaluation of the Wilson loop averages, unless 
the quenched approximation is adopted. Therefore, the averaging 
formulas that we shall encounter should be understood with the
presence of the quark field determinant. The expansions that 
will be used for the quark propagator in the external gluon field
can also be repeated inside the quark field determinant if
Wilson loop averages are to be evaluated.
\par
Using now for $S(A)$ Eq. (\rf{3e11}), one obtains:
\bea \lb{4e3}
& &S_{(n)}(x,x';y_{n-1},\ldots,y_1)=\frac{1}{N_c}S(x,x')
\langle U(x',y_{n-1})
\cdots U(y_1,x)U(x,x')\rangle\nonumber \\
& &\ \ +\frac{1}{N_c}\langle U(x',y_{n-1})\cdots U(y_1,x)
\Big(\frac{\bar\delta S(x,y_n)}{\bar\delta y_n^{\alpha +}}U(x,y_n)
+S(x,y_n)\frac{\bar\delta U(x,y_n)}{\bar\delta y_n^{\alpha -}}\Big)
\gamma^{\alpha}S(y_n,x';A)\rangle\nonumber \\
& &\ \ =S(x,x')\,e^{{\displaystyle F_{n+1}(x',y_{n-1},\ldots,y_1,x)}}
\nonumber \\
& &\ \ \ \ +\Big(\frac{\bar\delta S(x,y_n)}{\bar\delta y_n^{\alpha +}}
+S(x,y_n)\frac{\bar\delta }{\bar\delta y_n^{\alpha -}}\Big)
\,\gamma^{\alpha}\,S_{(n+1)}(y_n,x';y_{n-1},\ldots,y_1,x).
\eea
A graphical representation of this equation for $n=3$ is shown in 
Fig. \rf{4f1}.
\par
\bfg
\vspace*{0.5 cm}
\bc
\begin{picture}(0,0)%
\includegraphics{4f1.pstex}%
\end{picture}%
\setlength{\unitlength}{2763sp}%
\begingroup\makeatletter\ifx\SetFigFont\undefined%
\gdef\SetFigFont#1#2#3#4#5{%
  \reset@font\fontsize{#1}{#2pt}%
  \fontfamily{#3}\fontseries{#4}\fontshape{#5}%
  \selectfont}%
\fi\endgroup%
\begin{picture}(10136,2005)(751,-4385)
\put(6301,-3961){\makebox(0,0)[lb]{\smash{{\SetFigFont{12}{14.4}{\familydefault}{\mddefault}{\updefault}{\color[rgb]{0,0,0}$+$}%
}}}}
\put(9226,-3886){\makebox(0,0)[lb]{\smash{{\SetFigFont{12}{14.4}{\familydefault}{\mddefault}{\updefault}{\color[rgb]{0,0,0}$+$}%
}}}}
\put(2776,-4111){\makebox(0,0)[lb]{\smash{{\SetFigFont{10}{12.0}{\familydefault}{\mddefault}{\updefault}{\color[rgb]{0,0,0}$=$}%
}}}}
\put(5626,-4111){\makebox(0,0)[lb]{\smash{{\SetFigFont{10}{12.0}{\familydefault}{\mddefault}{\updefault}{\color[rgb]{0,0,0}$+$}%
}}}}
\put(2251,-4261){\makebox(0,0)[lb]{\smash{{\SetFigFont{8}{9.6}{\familydefault}{\mddefault}{\updefault}{\color[rgb]{0,0,0}$x'$}%
}}}}
\put(6826,-2536){\makebox(0,0)[lb]{\smash{{\SetFigFont{8}{9.6}{\familydefault}{\mddefault}{\updefault}{\color[rgb]{0,0,0}$y_1$}%
}}}}
\put(9676,-2536){\makebox(0,0)[lb]{\smash{{\SetFigFont{8}{9.6}{\familydefault}{\mddefault}{\updefault}{\color[rgb]{0,0,0}$y_1$}%
}}}}
\put(3451,-4261){\makebox(0,0)[lb]{\smash{{\SetFigFont{8}{9.6}{\familydefault}{\mddefault}{\updefault}{\color[rgb]{0,0,0}$x$}%
}}}}
\put(3751,-2911){\makebox(0,0)[lb]{\smash{{\SetFigFont{8}{9.6}{\familydefault}{\mddefault}{\updefault}{\color[rgb]{0,0,0}$y_1$}%
}}}}
\put(4276,-4336){\makebox(0,0)[lb]{\smash{{\SetFigFont{8}{9.6}{\familydefault}{\mddefault}{\updefault}{\color[rgb]{0,0,0}$S$}%
}}}}
\put(976,-2911){\makebox(0,0)[lb]{\smash{{\SetFigFont{8}{9.6}{\familydefault}{\mddefault}{\updefault}{\color[rgb]{0,0,0}$y_1$}%
}}}}
\put(2101,-2911){\makebox(0,0)[lb]{\smash{{\SetFigFont{8}{9.6}{\familydefault}{\mddefault}{\updefault}{\color[rgb]{0,0,0}$y_2$}%
}}}}
\put(5101,-4261){\makebox(0,0)[lb]{\smash{{\SetFigFont{8}{9.6}{\familydefault}{\mddefault}{\updefault}{\color[rgb]{0,0,0}$x'$}%
}}}}
\put(7801,-4186){\makebox(0,0)[lb]{\smash{{\SetFigFont{8}{9.6}{\familydefault}{\mddefault}{\updefault}{\color[rgb]{0,0,0}$x'$}%
}}}}
\put(751,-4261){\makebox(0,0)[lb]{\smash{{\SetFigFont{8}{9.6}{\familydefault}{\mddefault}{\updefault}{\color[rgb]{0,0,0}$x$}%
}}}}
\put(5926,-3436){\makebox(0,0)[lb]{\smash{{\SetFigFont{8}{9.6}{\familydefault}{\mddefault}{\updefault}{\color[rgb]{0,0,0}$S$}%
}}}}
\put(8701,-3436){\makebox(0,0)[lb]{\smash{{\SetFigFont{8}{9.6}{\familydefault}{\mddefault}{\updefault}{\color[rgb]{0,0,0}$S$}%
}}}}
\put(4876,-2911){\makebox(0,0)[lb]{\smash{{\SetFigFont{8}{9.6}{\familydefault}{\mddefault}{\updefault}{\color[rgb]{0,0,0}$y_2$}%
}}}}
\put(6976,-4186){\makebox(0,0)[lb]{\smash{{\SetFigFont{8}{9.6}{\familydefault}{\mddefault}{\updefault}{\color[rgb]{0,0,0}$S_{(4)}$}%
}}}}
\put(9751,-4186){\makebox(0,0)[lb]{\smash{{\SetFigFont{8}{9.6}{\familydefault}{\mddefault}{\updefault}{\color[rgb]{0,0,0}$S_{(4)}$}%
}}}}
\put(10576,-4186){\makebox(0,0)[lb]{\smash{{\SetFigFont{8}{9.6}{\familydefault}{\mddefault}{\updefault}{\color[rgb]{0,0,0}$x'$}%
}}}}
\put(1426,-4261){\makebox(0,0)[lb]{\smash{{\SetFigFont{8}{9.6}{\familydefault}{\mddefault}{\updefault}{\color[rgb]{0,0,0}$S_{(3)}$}%
}}}}
\put(6151,-3061){\makebox(0,0)[lb]{\smash{{\SetFigFont{8}{9.6}{\familydefault}{\mddefault}{\updefault}{\color[rgb]{0,0,0}$x$}%
}}}}
\put(8926,-3061){\makebox(0,0)[lb]{\smash{{\SetFigFont{8}{9.6}{\familydefault}{\mddefault}{\updefault}{\color[rgb]{0,0,0}$x$}%
}}}}
\put(8326,-4111){\makebox(0,0)[lb]{\smash{{\SetFigFont{10}{12.0}{\familydefault}{\mddefault}{\updefault}{\color[rgb]{0,0,0}$+$}%
}}}}
\put(7726,-2986){\makebox(0,0)[lb]{\smash{{\SetFigFont{8}{9.6}{\familydefault}{\mddefault}{\updefault}{\color[rgb]{0,0,0}$y_2$}%
}}}}
\put(10426,-2986){\makebox(0,0)[lb]{\smash{{\SetFigFont{8}{9.6}{\familydefault}{\mddefault}{\updefault}{\color[rgb]{0,0,0}$y_2$}%
}}}}
\put(6526,-4186){\makebox(0,0)[lb]{\smash{{\SetFigFont{8}{9.6}{\familydefault}{\mddefault}{\updefault}{\color[rgb]{0,0,0}$y_3$}%
}}}}
\put(9301,-4186){\makebox(0,0)[lb]{\smash{{\SetFigFont{8}{9.6}{\familydefault}{\mddefault}{\updefault}{\color[rgb]{0,0,0}$y_3$}%
}}}}
\put(4201,-3436){\makebox(0,0)[lb]{\smash{{\SetFigFont{8}{9.6}{\familydefault}{\mddefault}{\updefault}{\color[rgb]{0,0,0}$W_4$}%
}}}}
\end{picture}%

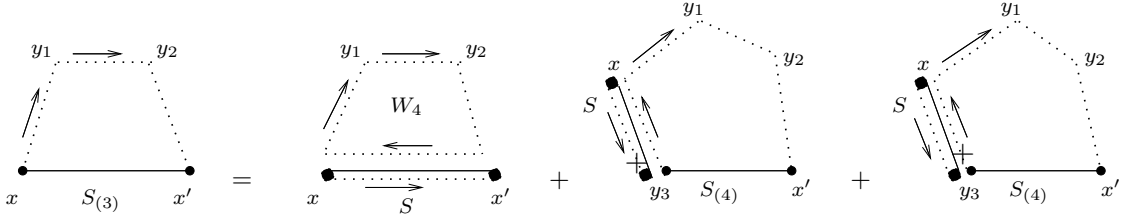
\caption{Graphical representation of Eq. (\rf{4e3}) for $n=3$. Same 
conventions as in Fig. \rf{3f1}. $y_3$ is an integration variable.}
\lb{4f1}
\ec
\efg
The Green's function $S_{(n)}$ satisfies the following equation of
motion with $x$:
\bea \lb{4e4}
& &(i\gamma.\partial_{(x)}-m)S_{(n)}(x,x';y_{n-1},\ldots,y_1)=
i\delta^4(x-x')e^{{\displaystyle F_{n}(x,y_{n-1},\ldots,y_1)}}
\nonumber \\
& &\ \ \ +i\gamma^{\mu}\frac{\bar\delta S_{(n)}(x,x';y_{n-1},\ldots,y_1)}
{\bar\delta x^{\mu -}}.
\eea
A graphical representation of this equation for $n=1$ and $n=3$ is 
shown in Fig. \rf{4f2}. 
\par
\bfg
\vspace*{0.5 cm}
\bc
\begin{picture}(0,0)%
\includegraphics{4f2.pstex}%
\end{picture}%
\setlength{\unitlength}{2960sp}%
\begingroup\makeatletter\ifx\SetFigFont\undefined%
\gdef\SetFigFont#1#2#3#4#5{%
  \reset@font\fontsize{#1}{#2pt}%
  \fontfamily{#3}\fontseries{#4}\fontshape{#5}%
  \selectfont}%
\fi\endgroup%
\begin{picture}(8186,3250)(1351,-4394)
\put(2626,-1636){\makebox(0,0)[lb]{\smash{{\SetFigFont{9}{10.8}{\familydefault}{\mddefault}{\updefault}{\color[rgb]{0,0,0}$x$}%
}}}}
\put(4126,-1636){\makebox(0,0)[lb]{\smash{{\SetFigFont{9}{10.8}{\familydefault}{\mddefault}{\updefault}{\color[rgb]{0,0,0}$x'$}%
}}}}
\put(6751,-1636){\makebox(0,0)[lb]{\smash{{\SetFigFont{9}{10.8}{\familydefault}{\mddefault}{\updefault}{\color[rgb]{0,0,0}$x$}%
}}}}
\put(8326,-1636){\makebox(0,0)[lb]{\smash{{\SetFigFont{9}{10.8}{\familydefault}{\mddefault}{\updefault}{\color[rgb]{0,0,0}$x'$}%
}}}}
\put(5476,-4261){\makebox(0,0)[lb]{\smash{{\SetFigFont{9}{10.8}{\familydefault}{\mddefault}{\updefault}{\color[rgb]{0,0,0}$i\delta^4(x-x')$}%
}}}}
\put(4276,-4261){\makebox(0,0)[lb]{\smash{{\SetFigFont{9}{10.8}{\familydefault}{\mddefault}{\updefault}{\color[rgb]{0,0,0}$x'$}%
}}}}
\put(3001,-2911){\makebox(0,0)[lb]{\smash{{\SetFigFont{9}{10.8}{\familydefault}{\mddefault}{\updefault}{\color[rgb]{0,0,0}$y_1$}%
}}}}
\put(4126,-2911){\makebox(0,0)[lb]{\smash{{\SetFigFont{9}{10.8}{\familydefault}{\mddefault}{\updefault}{\color[rgb]{0,0,0}$y_2$}%
}}}}
\put(2776,-4261){\makebox(0,0)[lb]{\smash{{\SetFigFont{9}{10.8}{\familydefault}{\mddefault}{\updefault}{\color[rgb]{0,0,0}$x$}%
}}}}
\put(3451,-4261){\makebox(0,0)[lb]{\smash{{\SetFigFont{9}{10.8}{\familydefault}{\mddefault}{\updefault}{\color[rgb]{0,0,0}$S_{(3)}$}%
}}}}
\put(9226,-4336){\makebox(0,0)[lb]{\smash{{\SetFigFont{9}{10.8}{\familydefault}{\mddefault}{\updefault}{\color[rgb]{0,0,0}$x'$}%
}}}}
\put(7951,-2986){\makebox(0,0)[lb]{\smash{{\SetFigFont{9}{10.8}{\familydefault}{\mddefault}{\updefault}{\color[rgb]{0,0,0}$y_1$}%
}}}}
\put(9076,-2986){\makebox(0,0)[lb]{\smash{{\SetFigFont{9}{10.8}{\familydefault}{\mddefault}{\updefault}{\color[rgb]{0,0,0}$y_2$}%
}}}}
\put(7726,-4336){\makebox(0,0)[lb]{\smash{{\SetFigFont{9}{10.8}{\familydefault}{\mddefault}{\updefault}{\color[rgb]{0,0,0}$x$}%
}}}}
\put(8401,-4336){\makebox(0,0)[lb]{\smash{{\SetFigFont{9}{10.8}{\familydefault}{\mddefault}{\updefault}{\color[rgb]{0,0,0}$S_{(3)}$}%
}}}}
\put(5326,-2911){\makebox(0,0)[lb]{\smash{{\SetFigFont{9}{10.8}{\familydefault}{\mddefault}{\updefault}{\color[rgb]{0,0,0}$y_1$}%
}}}}
\put(6526,-2911){\makebox(0,0)[lb]{\smash{{\SetFigFont{9}{10.8}{\familydefault}{\mddefault}{\updefault}{\color[rgb]{0,0,0}$y_2$}%
}}}}
\put(1351,-1336){\makebox(0,0)[lb]{\smash{{\SetFigFont{9}{10.8}{\familydefault}{\mddefault}{\updefault}{\color[rgb]{0,0,0}$(i\gamma.\partial_x - m)$}%
}}}}
\put(4651,-1411){\makebox(0,0)[lb]{\smash{{\SetFigFont{9}{10.8}{\familydefault}{\mddefault}{\updefault}{\color[rgb]{0,0,0}$=$}%
}}}}
\put(6001,-1411){\makebox(0,0)[lb]{\smash{{\SetFigFont{9}{10.8}{\familydefault}{\mddefault}{\updefault}{\color[rgb]{0,0,0}$+$}%
}}}}
\put(4876,-1636){\makebox(0,0)[lb]{\smash{{\SetFigFont{9}{10.8}{\familydefault}{\mddefault}{\updefault}{\color[rgb]{0,0,0}$i\delta^4(x-x')$}%
}}}}
\put(1426,-3961){\makebox(0,0)[lb]{\smash{{\SetFigFont{9}{10.8}{\familydefault}{\mddefault}{\updefault}{\color[rgb]{0,0,0}$(i\gamma.\partial_x - m)$}%
}}}}
\put(4951,-4036){\makebox(0,0)[lb]{\smash{{\SetFigFont{11}{13.2}{\familydefault}{\mddefault}{\updefault}{\color[rgb]{0,0,0}$=$}%
}}}}
\put(7051,-4036){\makebox(0,0)[lb]{\smash{{\SetFigFont{9}{10.8}{\familydefault}{\mddefault}{\updefault}{\color[rgb]{0,0,0}$+$}%
}}}}
\put(3301,-1636){\makebox(0,0)[lb]{\smash{{\SetFigFont{9}{10.8}{\familydefault}{\mddefault}{\updefault}{\color[rgb]{0,0,0}$S$}%
}}}}
\put(7426,-1636){\makebox(0,0)[lb]{\smash{{\SetFigFont{9}{10.8}{\familydefault}{\mddefault}{\updefault}{\color[rgb]{0,0,0}$S$}%
}}}}
\put(6901,-1336){\makebox(0,0)[lb]{\smash{{\SetFigFont{14}{16.8}{\familydefault}{\mddefault}{\updefault}{\color[rgb]{0,0,0}$\times$}%
}}}}
\put(7801,-3886){\makebox(0,0)[lb]{\smash{{\SetFigFont{14}{16.8}{\familydefault}{\mddefault}{\updefault}{\color[rgb]{0,0,0}$\times$}%
}}}}
\put(5851,-3361){\makebox(0,0)[lb]{\smash{{\SetFigFont{9}{10.8}{\familydefault}{\mddefault}{\updefault}{\color[rgb]{0,0,0}$W_3$}%
}}}}
\end{picture}%

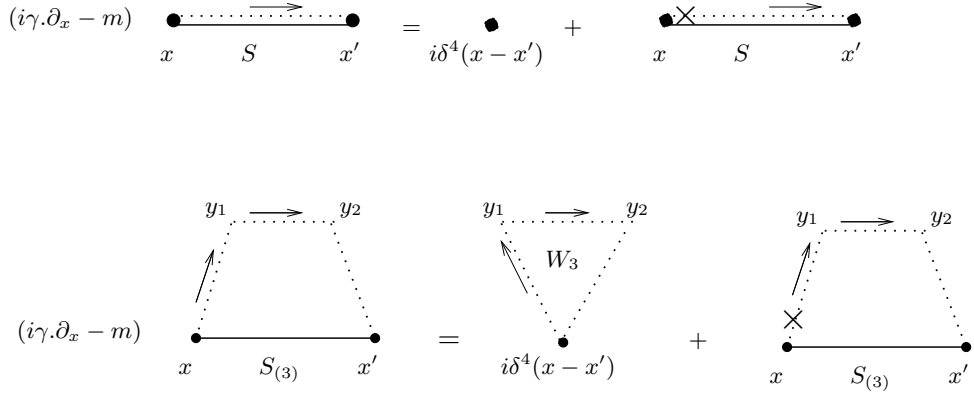
\caption{Graphical representation of the equations of motion of
$S_{(1)}\equiv S$ and $S_{(3)}$. Same conventions as in Fig. \rf{3f1}.}
\lb{4f2}
\ec
\efg
\par

\section{Integral equation} \lb{s5}
\setcounter{equation}{0}


The equations of motion of the gauge invariant Green's functions 
$S_{(n)}$ [Eqs. (\rf{3e8a}) and (\rf{4e4})] involve in their right-hand 
sides as unknowns the rigid path derivative of the Green's functions. 
The core of the problem amounts therefore to
the evaluation of the rigid path derivative of Green's functions.
That task, however, is facilitated by the functional relations
(\rf{4e3}), which relate two successive Green's functions
with increasing index. They allow the evaluation of the rigid path
derivative of a Green's function in terms of a similar derivative of 
a Wilson loop average and the derivative of a Green's function with
a higher index. Systematic repetition of this procedure allows us
therefore to express the rigid path derivative of a Green's function
in terms of a series of Green's functions whose coefficients are
functional derivatives of Wilson loop averages. One thus obtains
chains of coupled integral (or integro-differential) equations
between the various Green's functions. At the end, each Green's function
$S_{(n)}$ can be expressed, at leading order of an expansion, by
means of the functional relation (\rf{4e3}), in terms of the 
lowest-order Green's function $S$, and thus an equation where solely
the Green's function $S$ would appear becomes reachable.
\par
In the present work we are mainly interested by the simplest Green's 
function $S$ and therefore we shall concentrate our considerations 
on the equation of motion of that quantity.
\par
The rigid path derivative of $S_{(n)}$ along the segment $xy_1$ is
obtained from Eq. (\rf{4e3}):
\bea \lb{5e1}
& &\frac{\bar\delta S_{(n)}(x,x';y_{n-1},\ldots,y_1)}
{\bar\delta x^{\mu -}}=\frac{\bar\delta F_{n+1}}{\bar\delta x^{\mu -}}
\,e^{{\displaystyle F_{n+1}(x',y_{n-1},\ldots,y_1,x)}}\,S(x,x')
\nonumber \\
& &\ \ \ \ +\frac{\bar\delta}{\bar\delta x^{\mu -}}\,
\Big(\frac{\bar\delta S(x,y_n)}{\bar\delta y_n^{\alpha +}}
+S(x,y_n)\frac{\bar\delta }{\bar\delta y_n^{\alpha -}}\Big)
\,\gamma^{\alpha}\,S_{(n+1)}(y_n,x';y_{n-1},\ldots,y_1,x).
\eea
Eliminating in the right-hand side of the latter equation the
product $e^{F_{n+1}}S$ through Eq. (\rf{4e3}), one obtains the
equation
\bea \lb{5e2}
& &\frac{\bar\delta S_{(n)}(x,x';y_{n-1},\ldots,y_1)}
{\bar\delta x^{\mu -}}=\frac{\bar\delta F_{n+1}(x',y_{n-1},\ldots,y_1,x)}
{\bar\delta x^{\mu -}}\,S_{(n)}(x,x';y_{n-1},\ldots,y_1)\nonumber \\
& &\ \ \ \ +\Big(\frac{\bar\delta}{\bar\delta x^{\mu -}}-
\frac{\bar\delta F_{n+1}}{\bar\delta x^{\mu -}}\Big)\,
\Big(\frac{\bar\delta S(x,y_n)}{\bar\delta y_n^{\alpha +}}
+S(x,y_n)\frac{\bar\delta }{\bar\delta y_n^{\alpha -}}\Big)
\,\gamma^{\alpha}\,S_{(n+1)}(y_n,x';y_{n-1},\ldots,y_1,x).
\nonumber \\
& &
\eea
[Integrations on new variables in the right-hand sides are implicit.]
For $n=1$, one has:
\bea \lb{5e3}
& &\frac{\bar\delta S(x,x')}{\bar\delta x^{\mu -}}=
\frac{\bar\delta F_{2}(x',x)}{\bar\delta x^{\mu -}}\,S(x,x')
\nonumber \\
& &\ \ \ \ +\Big(\frac{\bar\delta}{\bar\delta x^{\mu -}}-
\frac{\bar\delta F_{2}(x',x)}{\bar\delta x^{\mu -}}\Big)\,
\Big(\frac{\bar\delta S(x,y_1)}{\bar\delta y_1^{\alpha_1 +}}
+S(x,y_1)\frac{\bar\delta}{\bar\delta y_1^{\alpha_1 -}}\Big)
\,\gamma^{\alpha_1}\,S_{(2)}(y_1,x';x).
\eea
\par
We next evaluate, in Eq. (\rf{5e3}), the action of the path derivation 
operators on $S_{(2)}$. The operator $\bar\delta/\bar\delta x^{\mu -}$
acts here on the segment $xx'$ and therefore it can be brought without 
harm to the utmost right, where an equation similar to Eq. (\rf{5e2}) 
(with a relabelling of some arguments, the point $x$ being now a junction
point on the path $y_1xx'$ of $S_{(2)}$) is used with $n=2$ and then 
$\bar\delta F_{3}/\bar\delta x^{\mu -}$ is brought back to the left;
during the last operation it is also submitted to the action of the
operator $\bar\delta/\bar\delta y^{\alpha_1 -}$. The resulting terms
that involve $S_{(2)}$ are:
\bea \lb{5e4}
& &\Big(\frac{\bar\delta F_3(x',x,y_1)}{\bar\delta x^{\mu -}}-
\frac{\bar\delta F_{2}(x',x)}{\bar\delta x^{\mu -}}\Big)\,
\Big(\frac{\bar\delta S(x,y_1)}{\bar\delta y_1^{\alpha_1 +}}
+S(x,y_1)\frac{\bar\delta}{\bar\delta y_1^{\alpha_1 -}}\Big)
\,\gamma^{\alpha_1}\,S_{(2)}(y_1,x';x)\nonumber \\
& &\ \ \ \ +\frac{\bar\delta^2 F_3(x',x,y_1)}{\bar\delta x^{\mu -}
\bar\delta y_1^{\alpha_1 -}}\,S(x,y_1)\,\gamma^{\alpha_1}\,
S_{(2)}(y_1,x';x).
\eea
Next, one observes that 
$\bar\delta S_{(2)}/\bar\delta y^{\alpha_1 -}$ is part of the equation 
of motion of $S_{(2)}$ [Eq. (\rf{4e4})], $y_1$ being now one of the
fermionic ends of $S_{(2)}$. Using the latter equation and making an 
integration by parts with respect to $y_1$, one arrives at a simplified 
expression
in which in the second derivative of $F_3$ the derivation
$\bar\delta/\bar\delta y_1^{\alpha_1 -}$, which is along the
segment $y_1x$, is replaced by the derivation 
$\bar\delta/\bar\delta y_1^{\alpha_1 +}$, which is along $y_1x'$.
[The delta functions of the equations of motion do not contribute
here because of the existence of the difference term 
$(\bar\delta F_3/\bar\delta x^{\mu -}-
\bar\delta F_{2}/\bar\delta x^{\mu -})$.] 
\par
The net result is, including also the resulting $S_{(3)}$ terms:
\bea \lb{5e5}
& &\frac{\bar\delta S(x,x')}{\bar\delta x^{\mu -}}=
\frac{\bar\delta F_{2}(x',x)}{\bar\delta x^{\mu -}}\,S(x,x')
-\frac{\bar\delta^2 F_3(x',x,y_1)}{\bar\delta x^{\mu -}
\bar\delta y_1^{\alpha_1 +}}\,S(x,y_1)\,\gamma^{\alpha_1}\,
S_{(2)}(y_1,x';x)
\nonumber \\
& &\ \ \ \ +\Big(\frac{\bar\delta S(x,y_1)}{\bar\delta y_1^{\alpha_1 +}}
+S(x,y_1)\frac{\bar\delta}{\bar\delta y_1^{\alpha_1 -}}\Big)
\,\gamma^{\alpha_1}\,
\Big(\frac{\bar\delta}{\bar\delta x^{\mu -}}-
\frac{\bar\delta F_{2}(x,x')}{\bar\delta x^{\mu -}}\Big)\nonumber\\
& &\ \ \ \ \ \ \ 
\times\Big(\frac{\bar\delta S(y_1,y_2)}{\bar\delta y_2^{\alpha_2 +}}
+S(y_1,y_2)\frac{\bar\delta}{\bar\delta y_2^{\alpha_2 -}}\Big)
\,\gamma^{\alpha_2}\,S_{(3)}(y_2,x';x,y_1).
\eea
\par
In the term containing $S_{(3)}$, the factors with the derivatives with
respect to $x$ and $y_2$ are treated in the same way as were those with 
$x$ and $y_1$ with $S_{(2)}$; they yield at the end the factor 
$\bar\delta^2 F_4/\bar\delta x^{\mu -}\bar\delta y_2^{\alpha_2 +}$ plus
a term with $S_{(4)}$ having a similar structure than the one with
$S_{(3)}$ above. Repeated use of the procedure described with $S_{(2)}$,
yields a series expansion in $S_{(n)}$ ($n=2,3,\ldots$) where all terms 
have similar structures. One obtains:
\bea \lb{5e6}
& &\frac{\bar\delta S(x,x')}{\bar\delta x^{\mu -}}=
\frac{\bar\delta F_{2}(x',x)}{\bar\delta x^{\mu -}}\,S(x,x')
-\frac{\bar\delta^2 F_3(x',x,y_1)}{\bar\delta x^{\mu -}
\bar\delta y_1^{\alpha_1 +}}\,S(x,y_1)\,\gamma^{\alpha_1}\,
S_{(2)}(y_1,x';x)
\nonumber \\
& &\  -\Big(\frac{\bar\delta S(x,y_1)}{\bar\delta y_1^{\alpha_1 +}}
+S(x,y_1)\frac{\bar\delta}{\bar\delta y_1^{\alpha_1 -}}\Big)
\,\gamma^{\alpha_1}\,\frac{\bar\delta^2 F_4(x',x,y_1,y_2)}
{\bar\delta x^{\mu -}\bar\delta y_2^{\alpha_2 +}}\,S(y_1,y_2)\,
\gamma^{\alpha_2}\,S_{(3)}(y_2,x';x,y_1)
\nonumber \\
& &\  -\sum_{n=4}^{\infty}\Big(\frac{\bar\delta S(x,y_1)}
{\bar\delta y_1^{\alpha_1 +}}+S(x,y_1)\frac{\bar\delta}
{\bar\delta y_1^{\alpha_1 -}}\Big)\,\gamma^{\alpha_1}\nonumber \\
& &\ \   
\times\cdots\times\,\Big(\frac{\bar\delta S(y_{n-3},y_{n-2})}
{\bar\delta y_{n-2}^{\alpha_{n-2} +}}+S(y_{n-3},y_{n-2})
\frac{\bar\delta}{\bar\delta y_{n-2}^{\alpha_{n-2} -}}\Big)
\gamma^{\alpha_{n-2}}\,\nonumber \\
& &\ \ \times\, 
\frac{\bar\delta^2 F_{n+1}(x',x,y_1,\ldots,y_{n-1})}{\bar\delta x^{\mu -}
\bar\delta y_{n-1}^{\alpha_{n-1} +}}\,S(y_{n-2},y_{n-1})\,
\gamma^{\alpha_{n-1}}\,S_{(n)}(y_{n-1},x';x,y_1,\ldots,y_{n-2}).
\eea
\par
Equation (\rf{5e6}) determines the action of the rigid path 
derivative on the Green's function. It is expressed in terms 
of derivatives of logarithms of Wilson loop averages appearing 
in a series of terms with skew-polygonal type contours. 
Although the resulting expression still contains other 
derivatives along internal lines, as well as implicit terms, 
their action will be determined with respect to the structure 
defined by Eq. (\rf{5e6}). The derivatives of the logarithms 
of Wilson loop averages, together with the accompanying quark
Green's function $S$, play here the role of kernels of the 
integral equation we are searching for; they are the analogs 
of the kernels made of propagators in Feynman diagrams appearing 
in Dyson-Schwinger equations. By analogy, we shall often call them 
diagrams.
\par
We now study the structure of the kernels that are present in the 
expansion (\rf{5e6}). We notice that in the term with 
$S_{(n)}$, the utmost left derivative related to $x$ is connected 
to the utmost right derivative related to $y_{n-1}$ through the
term $\bar\delta^2 F_{n+1}/\bar\delta x^-
\bar\delta y_{n-1}^+$; this does not leave room for the existence
of reducible type terms made of disjoint subsets of connnections;
such terms are actually parts of the definitions of the $S_{(n)}$s 
when expanded in terms of free propagators. 
In the present case, all remaining derivatives 
$\bar\delta/\bar\delta y_i$ ($i=1,\ldots,y_{n-2}$) either will act 
within the skew-polygonal line $xy_1\ldots y_{n-1}$ or will leave
that line to be connected to a larger contour associated with an 
$S_{(m)}$ with $m>n$. 
\par
To exhibit the latter feature, we consider in Eq. (\rf{5e6}) the 
general term $S_{(n)}$ and its associated derivative terms. We
notice that the derivatives 
$\bar\delta/\bar\delta y_i$ ($i=1,\ldots,y_{n-2}$) no longer act on
the fermionic end of $S_{(n)}$ (represented by $y_{n-1}$) but only
on the internal junction points of the skew-polygonal line    
$xy_1\ldots y_{n-1}$; hence, one cannot use the equation of 
motion of $S_{(n)}$ with these variables. We first consider the
action of $\bar\delta/\bar\delta y_{n-2}^-$. It is brought to the
right to act on $S_{(n)}$; during this operation it may also act
on the term $\bar\delta^2 F_{n+1}/\bar\delta x^-
\bar\delta y_{n-1}^+$ to yield a third-order derivative of 
$F_{n+1}$. When in front of $S_{(n)}$, it is convenient, to maintain
symmetry with $\bar\delta/\bar\delta y_{n-1}^+$, to replace 
$\bar\delta/\bar\delta y_{n-2}^-$ by 
$(\partial/\partial y_{n-2}-\bar\delta/\bar\delta y_{n-2}^+)$; the 
total derivative is then used for an integration by parts to 
convert $\bar\delta^3 F_{n+1}/\bar\delta x^-
\bar\delta y_{n-1}^+\bar\delta y_{n-2}^-$ into
$\bar\delta^3 F_{n+1}/\bar\delta x^-
\bar\delta y_{n-1}^+\bar\delta y_{n-2}^+$. The derivative 
$\bar\delta/\bar\delta y_{n-2}^+$ acting on $S_{(n)}$ has two kinds
of effect, according to Eq. (\rf{4e3}) (with a relabelling of variables):
in the first place, it yields the derivative term
$\bar\delta F_{n+1}/\bar\delta y_{n-2}^+$ and in the second, it
generates a new Green's function $S_{(n+1)}$ with the fermionic end
point $y_n$ (the global result is very similar to that of Eq. 
(\rf{5e2}) with a relabelling of variables). The term    
$\bar\delta F_{n+1}/\bar\delta y_{n-2}^+$ is an insertion along
the line $xy_1\ldots y_{n-1}$ and may also be submitted to other
derivations coming from the remaining variables $y_j$ ($j=1,\ldots,n-3$).
In the newly generated term with $S_{(n+1)}$, the actions of the
derivatives $\bar\delta/\bar\delta y_{n-2}^+$ and 
$\bar\delta/\bar\delta y_{n}^-$ can be combined to give
$\bar\delta^2 F_{n+2}/\bar\delta y_{n-2}^+\bar\delta y_n^+$. This term 
represents now a connection between the line $xy_1\ldots y_{n-1}$ and
the segment $y_{n-1}y_n$; it crosses the connection line $xy_{n-1}$
which had already appeared with the term   
$\bar\delta^2 F_{n+1}/\bar\delta x^-\bar\delta y_{n-1}^+$; the two
connections can therefore be interpreted as forming a crossed diagram. 
We observe that the appearance of a crossed diagram has been accompanied 
with the increase of the number of segments of the contour by one
unit. 
\par
The action of the remaining derivatives $\bar\delta/\bar\delta y_j^-$
($j=1,\ldots,n-3$) can be studied in a similar way. The qualitative
features displayed up to now remain unchanged. 
\par
We can summarize the above results by grouping the terms that appear in 
front of a Green's function $S_{(n)}$ into three categories: (i) A term 
that is completely connected: $F_{n+1}$ is submitted to the $n$ derivations
$\bar\delta^n/\bar\delta x^-\bar\delta y_1^+\ldots\bar\delta y_{n-1}^+$.
(ii) Crossed diagrams that involve at least one $F_{n+1}$ and some
other $F$s with lower indices. (iii) Nested diagrams, represented by
insertions within the connection line $xy_1\ldots y_{n-1}$ or within 
smaller connections of that line or within crossed diagrams. As a general 
property, no terms of reducible type (disjoint connections) exist.
\par
The general structure of the derivative $\bar\delta S/\bar\delta x^{\mu -}$
is:
\bea \lb{5e6a}
\frac{\bar\delta S(x,x')}{\bar\delta x^{\mu -}}&=&K_{1\mu -}(x',x)\,
S(x,x')+K_{2\mu -}(x',x,y_1)\,S_{(2)}(y_1,x';x)\nonumber \\
& &\ +\sum_{i=3}^{\infty}K_{i\mu -}(x',x,y_1,\ldots,y_{i-1})\,
S_{(i)}(y_{i-1},x';x,y_1,\ldots,y_{i-2}),
\eea
where the kernels $K_i$ ($i=1,\ldots,\infty$) are composed of the
three categories of terms quoted above and of ($i-1$) quark propagators 
$S$, and eventually of their derivatives, along the segments of the 
$(i+1)$-sided skew-polygons. The total number of derivatives contained
in $K_n$ is $n$.
\par
The explicit expression of $\bar\delta S/\bar\delta x^{\mu -}$ up to 
the fourth-order of its expansion is:
\par
\bea \lb{5e7} 
& &\frac{\bar\delta S(x,x')}{\bar\delta x^{\mu -}}=
\frac{\bar\delta F_{2}(x',x)}{\bar\delta x^{\mu -}}\,S(x,x')
-\frac{\bar\delta^2 F_3(x',x,y_1)}{\bar\delta x^{\mu -}
\bar\delta y_1^{\alpha_1 +}}\,S(x,y_1)\,\gamma^{\alpha_1}\,
S_{(2)}(y_1,x';x)
\nonumber \\
& &\ \ \ \ +\frac{\bar\delta^3 F_4(x',x,y_1,y_2)}{\bar\delta x^{\mu -}
\bar\delta y_1^{\alpha_1 +}\bar\delta y_2^{\alpha_2 +}}\,S(x,y_1)\,
\gamma^{\alpha_1}\,S(y_1,y_2)\,\gamma^{\alpha_2}\,S_{(3)}(y_2,x';x,y_1)
\nonumber\\
& &\ \ \ \ +\frac{\bar\delta^2 F_4}{\bar\delta x^{\mu -}
\bar\delta y_2^{\alpha_2 +}}\,S(x,y_1)\,\gamma^{\alpha_1}\, 
\Big(\frac{\bar\delta S(y_1,y_2)}{\bar\delta y_1^{\alpha_1 -}}
+S(y_1,y_2)\frac{\bar\delta F_4}{\bar\delta y_1^{\alpha_1 +}}\Big)\,
\gamma^{\alpha_2}\,S_{(3)}(y_2,x';x,y_1)\nonumber \\
& &\ \ \ \ -\bigg[\,\frac{\bar\delta^4 F_5(x',x,y_1,y_2,y_3)}
{\bar\delta x^{\mu -}\bar\delta y_1^{\alpha_1 +}
\bar\delta y_2^{\alpha_2 +}\bar\delta y_3^{\alpha_3 +}}\,+\,
\frac{\bar\delta^2 F_4(x',x,y_1,y_2)}
{\bar\delta x^{\mu -}\bar\delta y_2^{\alpha_2 +}}\,
\frac{\bar\delta^2 F_5(x',x,y_1,y_2,y_3)}
{\bar\delta y_1^{\alpha_1 +}\bar\delta y_3^{\alpha_3 +}}\,\bigg]
\nonumber \\
& &\ \ \ \ \ \ \times S(x,y_1)\,\gamma^{\alpha_1}\,S(y_1,y_2)\,
\gamma^{\alpha_2}\,S(y_2,y_3)\,\gamma^{\alpha_3}\,
S_{(4)}(y_3,x';x,y_1,y_2)\nonumber \\
& &\ \ \ \ -S(x,y_1)\,\gamma^{\alpha_1}\,\bigg[\,
\frac{\bar\delta^3 F_5}{\bar\delta x^{\mu -}\bar\delta y_2^{\alpha_2 +}
\bar\delta y_3^{\alpha_3 +}}\,\Big(\frac{\bar\delta S(y_1,y_2)}
{\bar\delta y_1^{\alpha_1 -}}+S(y_1,y_2)\frac{\bar\delta F_5}
{\bar\delta y_1^{\alpha_1 +}}\Big)\,\gamma^{\alpha_2}\,S(y_2,y_3)
\nonumber \\
& &\ \ \ \ \ \ +\frac{\bar\delta^3 F_5}{\bar\delta x^{\mu -}
\bar\delta y_1^{\alpha_1 +}\bar\delta y_3^{\alpha_3 +}}\,S(y_1,y_2)\,
\gamma^{\alpha_2}\,\Big(\frac{\bar\delta S(y_2,y_3)}
{\bar\delta y_2^{\alpha_2 -}}+S(y_2,y_3)\frac{\bar\delta F_5}
{\bar\delta y_2^{\alpha_2 +}}\Big)\nonumber \\
& &\ \ \ \ \ \ +\frac{\bar\delta^2 F_5}{\bar\delta x^{\mu -}
\bar\delta y_3^{\alpha_3 +}}\,\Big(\frac{\bar\delta S(y_1,y_2)}
{\bar\delta y_1^{\alpha_1 -}}+S(y_1,y_2)\frac{\bar\delta F_5}
{\bar\delta y_1^{\alpha_1 +}}\Big)\,\gamma^{\alpha_2}\,
\Big(\frac{\bar\delta S(y_2,y_3)}
{\bar\delta y_2^{\alpha_2 -}}+S(y_2,y_3)\frac{\bar\delta F_5}
{\bar\delta y_2^{\alpha_2 +}}\Big)\nonumber \\
& &\ \ \ \ \ \ +\frac{\bar\delta^2 F_5}{\bar\delta x^{\mu -}
\bar\delta y_3^{\alpha_3 +}}\,\frac{\bar\delta^2 F_5}
{\bar\delta y_1^{\alpha_1 +}\bar\delta y_2^{\alpha_2 +}}\,
S(y_1,y_2,)\,\gamma^{\alpha_2}\,S(y_2,y_3)\,\bigg]\,\gamma^{\alpha_3}\,
S_{(4)}(y_3,x';x,y_1,y_2)\nonumber \\
& &\ \ \ \ +\cdots\ \ .
\eea
The expansion, up to third-order terms (including $S_{(3)}$),
is represented graphically in Fig. \rf{5f1}. 
\par
\bfg
\vspace*{0.5 cm}
\bc
\begin{picture}(0,0)%
\includegraphics{5f1.pstex}%
\end{picture}%
\setlength{\unitlength}{2565sp}%
\begingroup\makeatletter\ifx\SetFigFont\undefined%
\gdef\SetFigFont#1#2#3#4#5{%
  \reset@font\fontsize{#1}{#2pt}%
  \fontfamily{#3}\fontseries{#4}\fontshape{#5}%
  \selectfont}%
\fi\endgroup%
\begin{picture}(9836,5989)(901,-6269)
\put(8551,-5386){\makebox(0,0)[lb]{\smash{{\SetFigFont{12}{14.4}{\familydefault}{\mddefault}{\updefault}{\color[rgb]{0,0,0}$\times$}%
}}}}
\put(3076,-1336){\makebox(0,0)[lb]{\smash{{\SetFigFont{8}{9.6}{\familydefault}{\mddefault}{\updefault}{\color[rgb]{0,0,0}$=$}%
}}}}
\put(3601,-1711){\makebox(0,0)[lb]{\smash{{\SetFigFont{8}{9.6}{\familydefault}{\mddefault}{\updefault}{\color[rgb]{0,0,0}$x$}%
}}}}
\put(5101,-1711){\makebox(0,0)[lb]{\smash{{\SetFigFont{8}{9.6}{\familydefault}{\mddefault}{\updefault}{\color[rgb]{0,0,0}$x'$}%
}}}}
\put(4276,-1711){\makebox(0,0)[lb]{\smash{{\SetFigFont{8}{9.6}{\familydefault}{\mddefault}{\updefault}{\color[rgb]{0,0,0}$S$}%
}}}}
\put(976,-1561){\makebox(0,0)[lb]{\smash{{\SetFigFont{8}{9.6}{\familydefault}{\mddefault}{\updefault}{\color[rgb]{0,0,0}$x$}%
}}}}
\put(2551,-1561){\makebox(0,0)[lb]{\smash{{\SetFigFont{8}{9.6}{\familydefault}{\mddefault}{\updefault}{\color[rgb]{0,0,0}$x'$}%
}}}}
\put(1651,-1561){\makebox(0,0)[lb]{\smash{{\SetFigFont{8}{9.6}{\familydefault}{\mddefault}{\updefault}{\color[rgb]{0,0,0}$S$}%
}}}}
\put(1126,-1261){\makebox(0,0)[lb]{\smash{{\SetFigFont{12}{14.4}{\familydefault}{\mddefault}{\updefault}{\color[rgb]{0,0,0}$\times$}%
}}}}
\put(5851,-1411){\makebox(0,0)[lb]{\smash{{\SetFigFont{8}{9.6}{\familydefault}{\mddefault}{\updefault}{\color[rgb]{0,0,0}$-$}%
}}}}
\put(6601,-1636){\makebox(0,0)[lb]{\smash{{\SetFigFont{12}{14.4}{\familydefault}{\mddefault}{\updefault}{\color[rgb]{0,0,0}$\times$}%
}}}}
\put(6451,-1861){\makebox(0,0)[lb]{\smash{{\SetFigFont{8}{9.6}{\familydefault}{\mddefault}{\updefault}{\color[rgb]{0,0,0}$x$}%
}}}}
\put(7726,-1861){\makebox(0,0)[lb]{\smash{{\SetFigFont{8}{9.6}{\familydefault}{\mddefault}{\updefault}{\color[rgb]{0,0,0}$x'$}%
}}}}
\put(8551,-1861){\makebox(0,0)[lb]{\smash{{\SetFigFont{8}{9.6}{\familydefault}{\mddefault}{\updefault}{\color[rgb]{0,0,0}$x$}%
}}}}
\put(9901,-1861){\makebox(0,0)[lb]{\smash{{\SetFigFont{8}{9.6}{\familydefault}{\mddefault}{\updefault}{\color[rgb]{0,0,0}$x'$}%
}}}}
\put(9076,-511){\makebox(0,0)[lb]{\smash{{\SetFigFont{8}{9.6}{\familydefault}{\mddefault}{\updefault}{\color[rgb]{0,0,0}$y_1$}%
}}}}
\put(6976,-1261){\makebox(0,0)[lb]{\smash{{\SetFigFont{8}{9.6}{\familydefault}{\mddefault}{\updefault}{\color[rgb]{0,0,0}$F_3$}%
}}}}
\put(9151,-1861){\makebox(0,0)[lb]{\smash{{\SetFigFont{8}{9.6}{\familydefault}{\mddefault}{\updefault}{\color[rgb]{0,0,0}$S_{(2)}$}%
}}}}
\put(4276,-886){\makebox(0,0)[lb]{\smash{{\SetFigFont{8}{9.6}{\familydefault}{\mddefault}{\updefault}{\color[rgb]{0,0,0}$F_2$}%
}}}}
\put(3826,-1186){\makebox(0,0)[lb]{\smash{{\SetFigFont{12}{14.4}{\familydefault}{\mddefault}{\updefault}{\color[rgb]{0,0,0}$\times$}%
}}}}
\put(8026,-1186){\makebox(0,0)[lb]{\smash{{\SetFigFont{12}{14.4}{\familydefault}{\mddefault}{\updefault}{\color[rgb]{0,0,0}$\times$}%
}}}}
\put(3301,-3961){\makebox(0,0)[lb]{\smash{{\SetFigFont{8}{9.6}{\familydefault}{\mddefault}{\updefault}{\color[rgb]{0,0,0}$x'$}%
}}}}
\put(1801,-3961){\makebox(0,0)[lb]{\smash{{\SetFigFont{8}{9.6}{\familydefault}{\mddefault}{\updefault}{\color[rgb]{0,0,0}$x$}%
}}}}
\put(2251,-2761){\makebox(0,0)[lb]{\smash{{\SetFigFont{12}{14.4}{\familydefault}{\mddefault}{\updefault}{\color[rgb]{0,0,0}$\times$}%
}}}}
\put(2101,-3736){\makebox(0,0)[lb]{\smash{{\SetFigFont{12}{14.4}{\familydefault}{\mddefault}{\updefault}{\color[rgb]{0,0,0}$\times$}%
}}}}
\put(2551,-3211){\makebox(0,0)[lb]{\smash{{\SetFigFont{8}{9.6}{\familydefault}{\mddefault}{\updefault}{\color[rgb]{0,0,0}$F_4$}%
}}}}
\put(1351,-3136){\makebox(0,0)[lb]{\smash{{\SetFigFont{8}{9.6}{\familydefault}{\mddefault}{\updefault}{\color[rgb]{0,0,0}$+$}%
}}}}
\put(3226,-6136){\makebox(0,0)[lb]{\smash{{\SetFigFont{8}{9.6}{\familydefault}{\mddefault}{\updefault}{\color[rgb]{0,0,0}$x'$}%
}}}}
\put(1726,-6136){\makebox(0,0)[lb]{\smash{{\SetFigFont{8}{9.6}{\familydefault}{\mddefault}{\updefault}{\color[rgb]{0,0,0}$x$}%
}}}}
\put(2026,-5911){\makebox(0,0)[lb]{\smash{{\SetFigFont{12}{14.4}{\familydefault}{\mddefault}{\updefault}{\color[rgb]{0,0,0}$\times$}%
}}}}
\put(2476,-5386){\makebox(0,0)[lb]{\smash{{\SetFigFont{8}{9.6}{\familydefault}{\mddefault}{\updefault}{\color[rgb]{0,0,0}$F_4$}%
}}}}
\put(5851,-6211){\makebox(0,0)[lb]{\smash{{\SetFigFont{8}{9.6}{\familydefault}{\mddefault}{\updefault}{\color[rgb]{0,0,0}$x'$}%
}}}}
\put(4351,-6211){\makebox(0,0)[lb]{\smash{{\SetFigFont{8}{9.6}{\familydefault}{\mddefault}{\updefault}{\color[rgb]{0,0,0}$x$}%
}}}}
\put(10351,-6211){\makebox(0,0)[lb]{\smash{{\SetFigFont{8}{9.6}{\familydefault}{\mddefault}{\updefault}{\color[rgb]{0,0,0}$x'$}%
}}}}
\put(8851,-6211){\makebox(0,0)[lb]{\smash{{\SetFigFont{8}{9.6}{\familydefault}{\mddefault}{\updefault}{\color[rgb]{0,0,0}$x$}%
}}}}
\put(9526,-6211){\makebox(0,0)[lb]{\smash{{\SetFigFont{8}{9.6}{\familydefault}{\mddefault}{\updefault}{\color[rgb]{0,0,0}$S_{(3)}$}%
}}}}
\put(10276,-4786){\makebox(0,0)[lb]{\smash{{\SetFigFont{8}{9.6}{\familydefault}{\mddefault}{\updefault}{\color[rgb]{0,0,0}$y_2$}%
}}}}
\put(7201,-961){\makebox(0,0)[lb]{\smash{{\SetFigFont{12}{14.4}{\familydefault}{\mddefault}{\updefault}{\color[rgb]{0,0,0}$\times$}%
}}}}
\put(3001,-2986){\makebox(0,0)[lb]{\smash{{\SetFigFont{12}{14.4}{\familydefault}{\mddefault}{\updefault}{\color[rgb]{0,0,0}$\times$}%
}}}}
\put(1576,-3436){\makebox(0,0)[lb]{\smash{{\SetFigFont{8}{9.6}{\familydefault}{\mddefault}{\updefault}{\color[rgb]{0,0,0}$S$}%
}}}}
\put(1801,-2461){\makebox(0,0)[lb]{\smash{{\SetFigFont{8}{9.6}{\familydefault}{\mddefault}{\updefault}{\color[rgb]{0,0,0}$y_1$}%
}}}}
\put(2401,-2311){\makebox(0,0)[lb]{\smash{{\SetFigFont{8}{9.6}{\familydefault}{\mddefault}{\updefault}{\color[rgb]{0,0,0}$S$}%
}}}}
\put(2401,-4561){\makebox(0,0)[lb]{\smash{{\SetFigFont{8}{9.6}{\familydefault}{\mddefault}{\updefault}{\color[rgb]{0,0,0}$S$}%
}}}}
\put(4951,-4636){\makebox(0,0)[lb]{\smash{{\SetFigFont{8}{9.6}{\familydefault}{\mddefault}{\updefault}{\color[rgb]{0,0,0}$S$}%
}}}}
\put(5026,-5461){\makebox(0,0)[lb]{\smash{{\SetFigFont{8}{9.6}{\familydefault}{\mddefault}{\updefault}{\color[rgb]{0,0,0}$F_4$}%
}}}}
\put(3601,-5386){\makebox(0,0)[lb]{\smash{{\SetFigFont{8}{9.6}{\familydefault}{\mddefault}{\updefault}{\color[rgb]{0,0,0}$+$}%
}}}}
\put(6076,-5461){\makebox(0,0)[lb]{\smash{{\SetFigFont{12}{14.4}{\familydefault}{\mddefault}{\updefault}{\color[rgb]{0,0,0}$\times$}%
}}}}
\put(901,-5311){\makebox(0,0)[lb]{\smash{{\SetFigFont{8}{9.6}{\familydefault}{\mddefault}{\updefault}{\color[rgb]{0,0,0}$+$}%
}}}}
\put(1276,-5386){\makebox(0,0)[lb]{\smash{{\SetFigFont{12}{14.4}{\familydefault}{\mddefault}{\updefault}{\color[rgb]{0,0,0}$\bigg($}%
}}}}
\put(4576,-5986){\makebox(0,0)[lb]{\smash{{\SetFigFont{12}{14.4}{\familydefault}{\mddefault}{\updefault}{\color[rgb]{0,0,0}$\times$}%
}}}}
\put(2251,-4786){\makebox(0,0)[lb]{\smash{{\SetFigFont{12}{14.4}{\familydefault}{\mddefault}{\updefault}{\color[rgb]{0,0,0}$\times$}%
}}}}
\put(5776,-4036){\makebox(0,0)[lb]{\smash{{\SetFigFont{8}{9.6}{\familydefault}{\mddefault}{\updefault}{\color[rgb]{0,0,0}$x'$}%
}}}}
\put(4276,-4036){\makebox(0,0)[lb]{\smash{{\SetFigFont{8}{9.6}{\familydefault}{\mddefault}{\updefault}{\color[rgb]{0,0,0}$x$}%
}}}}
\put(4951,-4036){\makebox(0,0)[lb]{\smash{{\SetFigFont{8}{9.6}{\familydefault}{\mddefault}{\updefault}{\color[rgb]{0,0,0}$S_{(3)}$}%
}}}}
\put(5701,-2686){\makebox(0,0)[lb]{\smash{{\SetFigFont{8}{9.6}{\familydefault}{\mddefault}{\updefault}{\color[rgb]{0,0,0}$y_2$}%
}}}}
\put(3751,-3286){\makebox(0,0)[lb]{\smash{{\SetFigFont{12}{14.4}{\familydefault}{\mddefault}{\updefault}{\color[rgb]{0,0,0}$\times$}%
}}}}
\put(8176,-5386){\makebox(0,0)[lb]{\smash{{\SetFigFont{12}{14.4}{\familydefault}{\mddefault}{\updefault}{\color[rgb]{0,0,0}$\bigg)$}%
}}}}
\put(5551,-5236){\makebox(0,0)[lb]{\smash{{\SetFigFont{12}{14.4}{\familydefault}{\mddefault}{\updefault}{\color[rgb]{0,0,0}$\times$}%
}}}}
\put(2926,-5161){\makebox(0,0)[lb]{\smash{{\SetFigFont{12}{14.4}{\familydefault}{\mddefault}{\updefault}{\color[rgb]{0,0,0}$\times$}%
}}}}
\put(4426,-4711){\makebox(0,0)[lb]{\smash{{\SetFigFont{8}{9.6}{\familydefault}{\mddefault}{\updefault}{\color[rgb]{0,0,0}$y_1$}%
}}}}
\put(1876,-4636){\makebox(0,0)[lb]{\smash{{\SetFigFont{8}{9.6}{\familydefault}{\mddefault}{\updefault}{\color[rgb]{0,0,0}$y_1$}%
}}}}
\put(3151,-2461){\makebox(0,0)[lb]{\smash{{\SetFigFont{8}{9.6}{\familydefault}{\mddefault}{\updefault}{\color[rgb]{0,0,0}$y_2$}%
}}}}
\put(3076,-4636){\makebox(0,0)[lb]{\smash{{\SetFigFont{8}{9.6}{\familydefault}{\mddefault}{\updefault}{\color[rgb]{0,0,0}$y_2$}%
}}}}
\put(9001,-4786){\makebox(0,0)[lb]{\smash{{\SetFigFont{8}{9.6}{\familydefault}{\mddefault}{\updefault}{\color[rgb]{0,0,0}$y_1$}%
}}}}
\put(5776,-4711){\makebox(0,0)[lb]{\smash{{\SetFigFont{8}{9.6}{\familydefault}{\mddefault}{\updefault}{\color[rgb]{0,0,0}$y_2$}%
}}}}
\put(7051,-436){\makebox(0,0)[lb]{\smash{{\SetFigFont{8}{9.6}{\familydefault}{\mddefault}{\updefault}{\color[rgb]{0,0,0}$y_1$}%
}}}}
\put(4426,-2686){\makebox(0,0)[lb]{\smash{{\SetFigFont{8}{9.6}{\familydefault}{\mddefault}{\updefault}{\color[rgb]{0,0,0}$y_1$}%
}}}}
\put(7951,-6136){\makebox(0,0)[lb]{\smash{{\SetFigFont{8}{9.6}{\familydefault}{\mddefault}{\updefault}{\color[rgb]{0,0,0}$x'$}%
}}}}
\put(6676,-4786){\makebox(0,0)[lb]{\smash{{\SetFigFont{8}{9.6}{\familydefault}{\mddefault}{\updefault}{\color[rgb]{0,0,0}$y_1$}%
}}}}
\put(7801,-4786){\makebox(0,0)[lb]{\smash{{\SetFigFont{8}{9.6}{\familydefault}{\mddefault}{\updefault}{\color[rgb]{0,0,0}$y_2$}%
}}}}
\put(6901,-4936){\makebox(0,0)[lb]{\smash{{\SetFigFont{12}{14.4}{\familydefault}{\mddefault}{\updefault}{\color[rgb]{0,0,0}$\times$}%
}}}}
\put(7201,-5386){\makebox(0,0)[lb]{\smash{{\SetFigFont{8}{9.6}{\familydefault}{\mddefault}{\updefault}{\color[rgb]{0,0,0}$F_4$}%
}}}}
\put(6451,-6136){\makebox(0,0)[lb]{\smash{{\SetFigFont{8}{9.6}{\familydefault}{\mddefault}{\updefault}{\color[rgb]{0,0,0}$x$}%
}}}}
\put(4051,-5686){\makebox(0,0)[lb]{\smash{{\SetFigFont{8}{9.6}{\familydefault}{\mddefault}{\updefault}{\color[rgb]{0,0,0}$S$}%
}}}}
\put(6601,-661){\makebox(0,0)[lb]{\smash{{\SetFigFont{8}{9.6}{\familydefault}{\mddefault}{\updefault}{\color[rgb]{0,0,0}$S$}%
}}}}
\put(1576,-5536){\makebox(0,0)[lb]{\smash{{\SetFigFont{8}{9.6}{\familydefault}{\mddefault}{\updefault}{\color[rgb]{0,0,0}$S$}%
}}}}
\end{picture}%

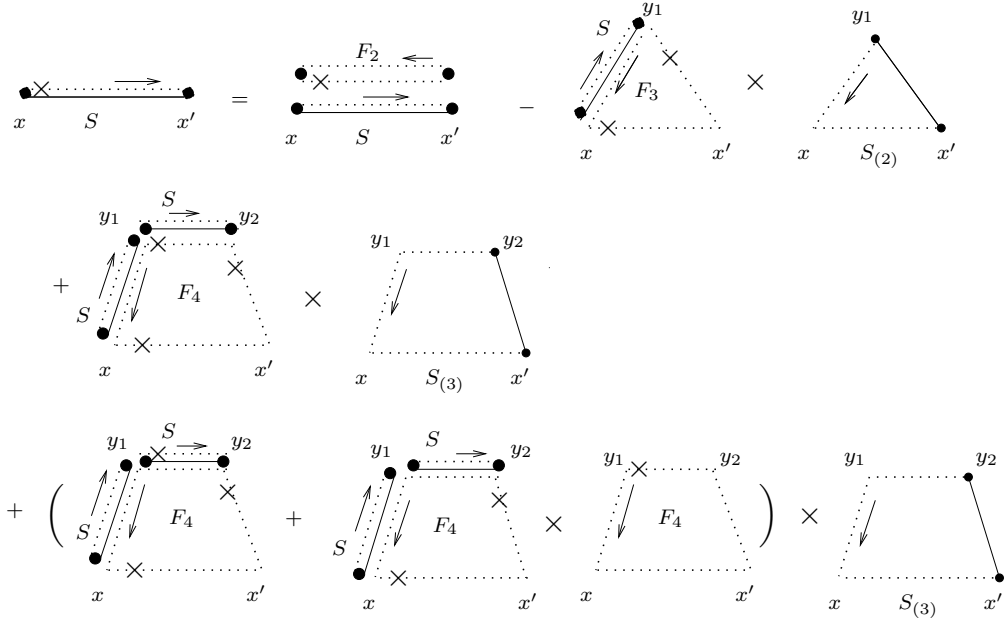
\caption{The expansion of $\bar\delta S/\bar\delta x^-$ up to third-order
terms. Same conventions as in Fig. \rf{3f1}. The $y$s are integration
variables.} 
\lb{5f1}
\ec
\efg
In Eq. (\rf{5e7}), the term in front of $S_{(2)}$ and the first terms
in front of $S_{(3)}$ and $S_{(4)}$ correspond to the completely
connected diagrams. The second term (within the first brackets) in front
of $S_{(4)}$ represents a crossed diagram. The remaining terms with
$S_{(3)}$ and $S_{(4)}$ correspond to nested diagrams.
\par
The persistence in the integral equation of nested diagrams may seem 
puzzling. In the Dyson--Schwinger equation for the self-energy, once
the internal propagators of diagrams are replaced by full propagators,
no nested diagram survives; this is equivalent to stating that the
summation of all internal diagrams in nested diagrams yields the
full propagators. In the present formalism, the survival of nested
diagrams is a consequence of a background effect induced by the 
Wilson loop; each sub-diagram is actually calculated in the presence of
the Wilson loop appearing at the order of the global diagram. At each 
order of the expansion the contour of the relevant Wilson loop changes 
with an increase of the number of segments forming the contour. The 
insertion of a low-order diagram in a higher-order diagram 
thus modifies the expression of the former, since it is now expressed 
with the new Wilson loop derivatives. In dealing with ordinary Feynman 
diagrams, one does not encounter the above background generated effects.
\par 
These statements can be explicitly checked in Eq. (\rf{5e7}). 
A way of ignoring background effects can proceed as follows:
(i) Assimilate an $n^{\mathrm{th}}$-order derivative of a function 
$F_m$ ($m\geq n+1$) to an ordianary $n$-point function with 
propagators attached to the end points of the corresponding segments, 
with the convention that a derivation of the type 
$\bar\delta/\bar\delta y^-$ can be 
converted into a derivation of the type $\bar\delta/\bar\delta y^+$ 
with a change of sign. (ii) Consider first the approximation where
$S_{(n)}$ is given by the first term of the expansion (\rf{4e3}).
(iii) When two loop contours corresponding to $W_i$ and $W_j$
have a common segment, replace the whole by a single contour
corresponding now to $W_{i+j-2}$.
It can then be verified that a systematic expansion of the terms 
$\bar\delta S/\bar\delta y^-$ inside the nested diagrams with an
iterative procedure cancels order-by-order all other terms
present in the nested diagrams and thus the latter disappear,
as expected, when no background effects are retained. At a second 
stage, considering the further terms of the expression of $S_{(n)}$
[Eq. (\rf{4e3})] one finds that the latter are themselves 
background generated effects and the repetition of the above 
procedure makes them in turn disappear.
\par
The rules of appearance of nested diagrams can be deduced from
Eq. (\rf{5e6}) and verified in Eq. (\rf{5e7}). 
\par
In two dimensions, where the Wilson loop averages are determined by 
the areas of the surfaces lying inside the contours \cite{kkk}, the
second functional derivatives of the functions $F$ are delta-functions 
and in general the nested diagrams disappear. One might expect that, 
in four dimensions, the nested diagrams, even if not disappearing, 
remain negligible on quantitative grounds. A more complete idea of 
their role could be obtained only when renormalization properties of 
the integral equation are studied.
\par
Equation (\rf{5e6}), together with relations (\rf{5e2}),
allows the calculation of the term
$\bar\delta S/\bar\delta x^-$ through an expansion involving an 
increasing number of functional derivatives of Wilson loops. 
The calculation of the expression of the kernel $K_n$ appearing
in the expansion (\rf{5e6a}) requires solely consideration of
terms of order lower or equal to $n$.
The integral form of the equation of motion (\rf{3e8a}) is:
\be \lb{5e8}
S(x,x')=S_0(x,x')+\int d^4x''\,S_0(x,x'')\,\gamma^{\mu}\,
\frac{\bar\delta S(x'',x')}{\bar\delta x^{''\mu -}},
\ee
in which one has to inject the expression of 
$\bar\delta S/\bar\delta x^-$ resulting from Eq. (\rf{5e6a}).
\par 
At short-distances, governed by perturbation theory, each 
derivation introduces a new power of the coupling constant and 
therefore the dominant terms in the expansion are the lowest-order 
ones. At large-distances, Wilson loops are saturated by the minimal 
surfaces having as supports the contours \cite{mm1,mm2,js}. Here 
also, the dominant contributions come from the lowest-order derivative 
terms. Therefore the expansion in Eq. (\rf{5e6a}) can be considered in 
general as a perturbative one whatever the distances are, provided that 
for each type of region the appropriate expressions are used for the 
Wilson loops. The first term of the expansion, represented by a single 
derivative, is null for symmetry reasons (the derivative 
$\bar\delta F_{2}(x',x)/\bar\delta x^-$ being orthogonal to $xx'$).
Hence the non-zero dominant term of the expansion is the second-order
derivative term. Furthermore, the various Green's functions $S_{(n)}$
are themselves dominated by their lowest-order expression of Eq.
(\rf{4e3}), involving only $S$ and a Wilson loop. In that approximation,
$\bar\delta S(x,x')/\bar\delta x^-$ takes the form
\be \lb{5e9}
\frac{\bar\delta S(x,x')}{\bar\delta x^{\mu -}}\simeq
-\int d^4y_1\,\frac{\bar\delta^2 F_3(x',x,y_1)}{\bar\delta x^{\mu -}
\bar\delta y_1^{\alpha_1 +}}\,
e^{{\displaystyle F_3(x',x,y_1)}}\,S(x,y_1)\,\gamma^{\alpha_1}\,
S(y_1,x').
\ee
Thus, the dominant part of the integral equation (\rf{5e8}) to be 
solved reduces to a closed form involving only the full propagator
$S$, the free propagator $S_0$, a Wilson loop average and its 
second-order rigid path derivative.
\par
The lack of manifest symmetry in Eq. (\rf{5e8}) between the 
coordinates $x$ and $x'$ is due to the presence of the closed
contours of the Wilson loops, which do not allow immediate
factorization of propagators through convolution operations.
Nevertheless, because of translation invariance, the Green's
function $S(x,x')$ depends only on the difference $(x-x')$;
therefore, once the integrations in Eq. (\rf{5e8}) are done, 
one should recover the desired symmetry. If, instead of the
equation of motion (\rf{3e8a}), relative to $x$, we had used 
the equation of motion (\rf{3e8b}), relative to $x'$, we would have
found an integral equation where $S_0(x'',x')$ acts on 
$\bar\delta S(x,x'')/\bar\delta x^{''\mu +}\gamma^{\mu}$ from the 
right.
\par
As a complementary remark with respect to the method of approach
developed in this work, we point out that another way of proceeding
would consist of trying to sum diagrams constructed with free quark
propagators associated with phase factors. This method actually 
corresponds to the use of the expansion equations (\rf{3e4}) and 
(\rf{3e5}), instead of (\rf{3e10}) and (\rf{3e11}). It, however,
becomes rapidly intricate due to the presence of the background 
Wilson loop effects. Nevertheless, the first few terms of 
Eqs. (\rf{5e6a}) and (\rf{5e7}) can be reconstituted rather easily.
The main aspects of this method are presented in Appendix \rf{ap1}.
\par
The integral equation (\rf{5e8}), together with the expressions
(\rf{5e6a}) or (\rf{5e7}), did not make any explicit reference to 
the quark self-energy function. The latter can be obtained once
the Green's function $S$ is calculated, through its inverse.
It is, however, also possible to construct it directly, by 
setting up a specific equation for it. This is presented in
Appendix \rf{ap2}.
\par       
The problem of solving Eq. (\rf{3e8a}), or its equivalent Eq. 
(\rf{5e8}), together with expression (\rf{5e6a}), leads us to the 
question of representation of the gauge invariant two-point Green's 
function. This will be considered in Sec. \rf{s6}.
\par

\section{Analyticity properties of the Green's function} \lb{s6}
\setcounter{equation}{0}


One of the advantages of paths along straight lines is the fact 
that the expressions of the corresponding Green's functions become 
dependent only on the end points of the paths. This feature in turn 
allows a simple transition to momentum space by Fourier transformation.
Much of the informations on Green's functions are provided from
momentum space, since it is there that their spectral properties 
are determined.
\par
From this point of view, the quark two-point gauge invariant
Green's functions hold a particular position. Because of confinement
of colored objects, it is not possible to cut the path joining
the quark to the antiquark by inserting in it a complete set of 
physical states, which are color singlets. This feature seems to 
suggest that gauge invariant two-point Green's functions should not 
have any singularities.
\par
The situation is, however, more complex than it seems. 
Gauge invariant two-point Green's functions possess singularities
originated from perturbation theory. This is corroborated by the 
integral equation (\rf{5e8}), in which the presence of the free 
quark propagator generates new singularities in the complete solution.
An analysis, starting from perturbation theory, is therefore
necessary.
\par
We shall admit that, in a domain where perturbation theory is 
valid, it is meaningful to consider quarks and gluons as
physical particles with positive energies, described by 
corresponding physical states. It is then
advantageous to consider the path-ordered phase factor (\rf{2e1})
in its representation given by the series expansion in terms of
the coupling constant $g$, the $n^{\mathrm{th}}$-order term of the
expansion containing $(n-1)$ gluon fields ($n\geq 1$). 
\par
Adopting here an operator formalism, we observe that the gauge 
invariant quark two-point function involves two kinds of orderings 
for its defining fields. The first is the path-ordering (or $P$-ordering)
which concerns the color index arrangements of the gluon fields 
according to their positions on the path. The second is the 
time-ordering ($T$-ordering) or chronological product which enters
in the definitions of Green's functions and operates once the
$P$-ordering is done. 
\par
Another advantage of paths along straight lines is that once the 
timelike or spacelike nature of the distance between the quark and 
the antiquark is fixed, the nature of the mutual distances of the 
gluon fields in the Green's function $S$ [Eq. (\rf{2e16})] is also 
fixed in the same way, because of their alignement along the segment 
joining the quark to the antiquark. Therefore, the chronological 
product of the $n^{\mathrm{th}}$-order terms in $S$ reduces to two 
terms, defined by the relative time between the quark and the antiquark. 
According to the definitions (\rf{2e1}) and (\rf{2e16}), for timelike
$(x'-x)$, if $(x^{'0}-x^0)>0$ the $T$-ordering will coincide with 
the $P$-ordering, while if $(x^{'0}-x^0)<0$ the $T$-ordering will 
be the opposite of the $P$-ordering (with a change of sign for the 
fermion fields), the color indices being already fixed from the
$P$-ordering. We are in a situation which is very similar to the 
case of the ordinary two-point function, with the difference that for 
an $n^{\mathrm{th}}$-order term there are $(n+1)$ fields instead of 
two ($(n-1)$ gluon, one quark and one antiquark fields). 
\par
Using for each of the two products which make the $T$-product
the spectral analysis with intermediate states, taking into account 
the bounds on the parameters of the $P$-ordering and using causality, 
one arrives at a generalized form of the K\"all\'en--Lehmann
representation for the Green's function $S$ in momentum space, in
which the cut starts on the real axis from the quark mass squared 
$m^2$ and extends to infinity \cite{kl,l,wght,schwb,thvth}. 
The generalization is due to the fact that each gluon field is 
integrated along the path and this introduces, when using for the 
latter a dimensionless parameter $\lambda$ varying between 0 and 1, 
a multiplicative factor $(x'-x)$, which is converted in momentum space 
into a derivation operator; each such factor increases by one unit 
the power of the denominator of the dispersion integral. Finally, 
because of the fact that we are dealing here with a gauge invariant 
quantity, we expect not to encounter at the end spurious infrared 
divergences. 
\par
To summarize the above results, we introduce the Fourier transform
of the Green's function $S(x,x')$, for which we also take into
account translation invariance:  
\be \lb{6e1}
S(x,x')=S(x-x')=\int \frac{d^4p}{(2\pi)^4}\,
e^{{\displaystyle -ip.(x-x')}}\,S(p).
\ee 
$S(p)$ has the following representation in terms of real spectral 
functions $\rho_1^{(n)}$ and $\rho_0^{(n)}$ ($n=1,\ldots,\infty$):
\be \lb{6e2}
S(p)=i\int_0^{\infty}ds'\,\sum_{n=1}^{\infty}\,
\frac{\big[\,\gamma.p\,\rho_1^{(n)}(s')+\rho_0^{(n)}(s')\,\big]}
{(p^2-s'+i\varepsilon)^n}.
\ee
This is a conservative representation of the various contributions
encountered above; simplifications or recombinations into more 
compact forms might still occur. It is evident that formally the 
powers of the denominators can be lowered by integration by parts; 
however, possible singularities of the spectral functions at 
threshold could prevent such an operation.
\par
We assume that the above representation, obtained from the domain of
perturbation theory, remains also valid in non-perturbative regimes.
One expects that the resulting singularities are strong enough 
to screen the quark pole and other physical type singularities.
\par
Further study is needed to define more accurately the properties
of the spectral functions. Nevertheless, representation (\rf{6e2}), 
or another one related to it, might be tried for the investigation 
of the solutions of the corresponding integral equation.
\par

\section{Summary and comments} \lb{s7}
\setcounter{equation}{0}


We have expressed the equation of motion of the gauge invariant
quark two-point function having a straight line path as an
integral or integro-differential equation involving the series of
all two-point functions with paths of skew-polygonal type, in
which the kernels are given by quark Green's functions and 
rigid path derivatives of the logarithms of the Wilson loop averages 
with contours made of these lines.
\par
Gauge invariant quark Green's functions also satisfy, in addition to 
their equations of motion related to the quark ends, other
equations of motion resulting from local deformations of their paths.
The latter equations are typically those of the path-ordered phase
factors and once Wilson loops are introduced through the calculations,
they reduce to the characteristic equations of Wilson loops, i.e.,
to the Bianchi identity and to the loop equation or Makeenko--Migdal
equation \cite{p,mm1,mm2}. We have not insisted on that aspect of
the problem, since it has been widely studied in the literature.
This implies that when the Wilson loop averages are used in the 
kernels of the above integral equations, they are understood as being
solutions (at least approximately) of their own equations of motion.
\par
We have emphasized the fact that the series of kernels appearing in 
the integral equation can be considered, on quantitative grounds, as
a perturbation series simultaneously for short- and large-distances
and therefore could be approximated, for a starting calculation,
by its lowest-order non-vanishing term. 
\par
A question which was not considered in the present work concerns
the renormalization properties of the gauge invariant Green's 
functions. These seem to be intimately related, through the 
integral equation, to those of the Wilson loop averages \cite{dv,bnsg}
and could be dealt with only when explicit expressions of the latter
are introduced.
\par
The method of functional relations between two-point Green's functions
with different numbers of segments along their paths, can also be
applied, with appropriate generalizations, to other $n$-point
Green's functions, with $n\geq 4$. That problem is particularly relevant 
for the derivation of a bound state equation for a quark-antiquark 
system.
\par
\vspace{0.25 cm}
\noindent
\textbf{Acknowledgements}
\par
This work was supported in part by the EU network FLAVIANET under 
Contract No. MRTN-CT-2006-035482.
\par

\appendix
\renewcommand{\theequation}{\Alph{section}.\arabic{equation}}

\section{Summation method} \lb{ap1}
\setcounter{equation}{0}

Integral equations of ordinary Green's functions represent 
in general the result of summing the classes of reducible 
diagrams in terms of free propagators. 
One should expect that a similar procedure might also be
operative in the case of gauge invariant Green's functions. 
\par 
To implement this method of approach, we should start with
expressions involving free quark propagators. To this end,
we have to use either of the two representations (\rf{3e4}) or
(\rf{3e5}) of the quark propagator in external field and expand 
$S(A)$ in the defining equation of $S$ [Eq. (\rf{4e2})]. 
Each of the two representations has its own advantages and 
could be preferred for a definite aim. Thus, representation
(\rf{3e4}) is more appropriate to obtain rapidly the structure
resulting from equations of motion with respect to $x$, while 
representation (\rf{3e5}) is more appropriate for the 
calculation of $\bar\delta S(x,x')/\bar\delta x^-$.
\par
Using first representation (\rf{3e4}), the expansion of $S(A)$ in
Eq. (\rf{4e2}) generates Wilson loops with skew-polygonal type 
contours, accompanied with free quark propagators:
\bea \lb{ae1}
& &S(x,x')=S_0(x,x')\,e^{{\displaystyle F_2(x',x)}}-
S_0(x,y_1)\,\gamma^{\alpha_1}\,S_0(y_1,x')\,
\frac{\bar\delta}{\bar\delta y_1^{\alpha_1+}}
\,e^{{\displaystyle F_3(x',x,y_1)}}\nonumber \\
& &\ \ \ \ +\sum_{j=2}^{\infty}(-1)^j\,S_0(x,y_1)\,\gamma^{\alpha_1}\,
S_0(y_1,y_2)\,\gamma^{\alpha_2}\,\cdots\,\gamma^{\alpha_j}\,
S_0(y_j,x')\nonumber \\
& &\ \ \ \ \ \ \ \ \times\frac{\bar\delta}{\bar\delta y_1^{\alpha_1+}}
\frac{\bar\delta}{\bar\delta y_2^{\alpha_2+}}\cdots
\frac{\bar\delta}{\bar\delta y_j^{\alpha_j+}}
\,e^{{\displaystyle F_{j+2}(x',x,y_1,\ldots,y_j)}}.
\eea
\par
A similar expansion can also be done for $S_{(n)}$ ($n>1$), starting 
from Eq. (\rf{4e1}):
\bea \lb{ae2}
& &S_{(n)}(x,x';z_{n-1},\ldots,z_1)=S_0(x,x')\,
e^{{\displaystyle F_{n+1}(x',z_{n-1},\ldots,z_1,x)}}\nonumber \\
& &\ \ \ \ -S_0(x,y_1)\,\gamma^{\alpha_1}\,S_0(y_1,x')\,
\frac{\bar\delta}{\bar\delta y_1^{\alpha_1+}}
\,e^{{\displaystyle F_{n+2}(x',z_{n-1},\ldots,z_1,x,y_1)}}\nonumber \\
& &\ \ \ \ +\sum_{j=2}^{\infty}(-1)^j\,S_0(x,y_1)\,\gamma^{\alpha_1}\,
S_0(y_1,y_2)\,\gamma^{\alpha_2}\,\cdots\,\gamma^{\alpha_j}\,
S_0(y_j,x')\nonumber \\
& &\ \ \ \ \ \ \ \ \times\frac{\bar\delta}{\bar\delta y_1^{\alpha_1+}}
\frac{\bar\delta}{\bar\delta y_2^{\alpha_2+}}\cdots
\frac{\bar\delta}{\bar\delta y_j^{\alpha_j+}}
\,e^{{\displaystyle F_{n+j+1}(x',z_{n-1},\ldots,z_1,x,y_1,\ldots,y_j)}}.
\eea
\par
Use of representation (\rf{3e5}) yields equivalent expressions for
$S$ and $S_{(n)}$ ($n>1$):
\bea \lb{ae3}
& &S(x,x')=\sum_{j=0}^{\infty}\,S_0(x,y_1)\,\gamma^{\alpha_1}\,
S_0(y_1,y_2)\,\gamma^{\alpha_2}\,\cdots\,\gamma^{\alpha_j}\,
S_0(y^j,x')\nonumber \\
& &\ \ \ \ \ \ \ \ \times\frac{\bar\delta}{\bar\delta y_1^{\alpha_1-}}
\frac{\bar\delta}{\bar\delta y_2^{\alpha_2-}}\cdots
\frac{\bar\delta}{\bar\delta y_j^{\alpha_j-}}
\,e^{{\displaystyle F_{j+2}(x',x,y_1,\ldots,y_j)}},
\eea
\bea \lb{ae4}
& &S_{(n)}(x,x';z_{n-1},\ldots,z_1)=\sum_{j=0}^{\infty}\,S_0(x,y_1)
\,\gamma^{\alpha_1}\,S_0(y_1,y_2)\,\gamma^{\alpha_2}\,\cdots\,
\gamma^{\alpha_j}\,S_0(y^j,x')\nonumber \\
& &\ \ \ \ \ \ \ \ \times\frac{\bar\delta}{\bar\delta y_1^{\alpha_1-}}
\frac{\bar\delta}{\bar\delta y_2^{\alpha_2-}}\cdots
\frac{\bar\delta}{\bar\delta y_j^{\alpha_j-}}
\,e^{{\displaystyle F_{n+j+1}(x',z_{n-1},\ldots,z_1,x,y_1,\ldots,y_j)}}.
\eea
Equations (\rf{ae3}) and (\rf{ae4}) could also have been obtained from
Eqs. (\rf{ae1}) and (\rf{ae2}), respectively, by integrations by parts;
at the internal junction points $y_i$ of the segments of the paths,
one has the equivalence relations 
$\partial/\partial y_i^{}=\bar\delta/\bar\delta y_i^+
+\bar\delta/\bar\delta y_i^-$ [Eq. (\rf{2e8})].
\par
The equations of motion with respect to $x$ can be evaluated easily
from representations (\rf{ae1}) and (\rf{ae2}). Because of the 
appearance of delta-functions from the propagators $S_0(x,y_1)$, there
are cancellations between successive terms and one finds:
\bea \lb{ae5}
& &(i\gamma.\partial_{(x)} -m)S(x,x')=i\delta^4(x-x')\nonumber \\
& &\ \ \ \ \ \ \ +\sum_{j=0}^{\infty}(-1)^j\,i\gamma^{\mu}\,
S_0(x,y_1)\,\gamma^{\alpha_1}\,S_0(y_1,y_2)\,\gamma^{\alpha_2}\,
\cdots\,\gamma^{\alpha_j}\,S_0(y_j,x')\nonumber \\
& &\ \ \ \ \ \ \ \times\frac{\bar\delta}{\bar\delta x^{\mu-}}
\frac{\bar\delta}{\bar\delta y_1^{\alpha_1+}}
\frac{\bar\delta}{\bar\delta y_2^{\alpha_2+}}\cdots
\frac{\bar\delta}{\bar\delta y_j^{\alpha_j+}}
\,e^{{\displaystyle F_{j+2}(x',x,y_1,\ldots,y_j)}}\nonumber \\
& &\ \ \ \ \ =i\delta^4(x-x')+i\gamma^{\mu}\,
\frac{\bar\delta S(x,x')}{\bar\delta x^{\mu-}},
\eea
\bea \lb{ae6}
& &(i\gamma.\partial_{(x)} -m)S_{(n)}(x,x';z_{n-1},\ldots,z_1)=
i\delta^4(x-x')\,e^{{\displaystyle F_{n}(x,z_{n-1},\ldots,z_1)}}
\nonumber \\
& &\ \ \ \ \ \ \ +\sum_{j=0}^{\infty}(-1)^j\,i\gamma^{\mu}\,
S_0(x,y_1)\,\gamma^{\alpha_1}\,S_0(y_1,y_2)\,\gamma^{\alpha_2}\,
\cdots\,\gamma^{\alpha_j}\,S_0(y_j,x')\nonumber \\
& &\ \ \ \ \ \ \ \times\frac{\bar\delta}{\bar\delta x^{\mu-}}
\frac{\bar\delta}{\bar\delta y_1^{\alpha_1+}}
\frac{\bar\delta}{\bar\delta y_2^{\alpha_2+}}\cdots
\frac{\bar\delta}{\bar\delta y_j^{\alpha_j+}}
\,e^{{\displaystyle F_{n+j+1}(x',z_{n-1},\ldots,z_1,x,y_1,\ldots,y_j)}}
\nonumber \\
& &\ \ \ \ \ =i\delta^4(x-x')\,e^{{\displaystyle F_{n}(x,z_{n-1},
\ldots,z_1)}}+i\gamma^{\mu}\,\frac{\bar\delta S_{(n)}(x,x';z_{n-1},
\ldots,z_1)}{\bar\delta x^{\mu-}}.
\eea
\par
Considering representation (\rf{ae3}), one immediately checks that
it has the structure of the integral equation (\rf{5e8}); this 
represents of course the integrated form of the equation of motion
(\rf{ae5}).
\par 
The action of the rigid path derivatives on the exponential 
functionals in Eqs. (\rf{ae1})-(\rf{ae4}) can be evaluated easily.
The aim is then to group the various terms that appear in the 
expression of $\bar\delta S(x,x')/\bar\delta x^-$ to bring the
latter into the form of Eq. (\rf{5e7}). We shall do this in a 
perturbative expansion with respect to the number of derivations, by
retaining up to derivatives of third order and showing that Eq. 
(\rf{5e7}) can be obtained up to the $S_{(3)}$ terms. That 
approximation is sufficient to illustrate the various aspects of the
method under consideration.
\par
For the calculation of $\bar\delta S(x,x')/\bar\delta x^-$, it is
preferable to start with representations (\rf{ae3}) and (\rf{ae4}) 
of $S$ and $S_{(n)}$; we shall indicate below the specific 
differences one meets when starting with representations (\rf{ae1})
and (\rf{ae2}), although the final result is the same.
The advantage of the former representations is that when a derivation
$\bar\delta/\bar\delta y_i^-$ acts on $S_{(i+1)}$ in which $y_i$ is a
fermionic end, it can be directly replaced in terms of the 
equation of motion operator and a delta-function, allowing an
integration by parts. This is not the case with the operator
$\bar\delta/\bar\delta y_i^+$, which first should be transformed into
$\bar\delta/\bar\delta y_i^-$ before an integration by parts be
possible. In the expansions (\rf{ae3}) and (\rf{ae4}), $S$ will be 
approximated with the first three terms, $S_{(2)}$ with the first two
and $S_{(3)}$ with the first term.
We shall also assume the backtracking property of the path-ordered
phase factors \cite{mk}, which means that $U(y,x)U(x,y)=1$, the same
path being run forth and back. This means that $W_2=1$ and $F_2=0$.
\par
Calculating $\bar\delta S(x,x')/\bar\delta x^-$ from Eq. (\rf{ae3})
one obtains:
\bea \lb{ae7}
& &\frac{\bar\delta S(x,x')}{\bar\delta x^{\mu -}}=
S_0(x,x')\,\frac{\bar\delta}{\bar\delta x^{\mu -}}
e^{{\displaystyle F_2(x',x)}}+
S_0(x,y_1)\gamma^{\alpha_1}S_0(y_1,x')\,
\frac{\bar\delta}{\bar\delta x^{\mu -}}
\frac{\bar\delta}{\bar\delta y_1^{\alpha_1-}}
\,e^{{\displaystyle F_3(x',x,y_1)}}\nonumber \\
& &\ \ \ \ +S_0(x,y_1)\,\gamma^{\alpha_1}\,S_0(y_1,y_2)\,
\gamma^{\alpha_2}\,S_0(y_2,x')\,
\frac{\bar\delta}{\bar\delta x^{\mu -}}
\frac{\bar\delta}{\bar\delta y_1^{\alpha_1-}}
\frac{\bar\delta}{\bar\delta y_2^{\alpha_2-}}
\,e^{{\displaystyle F_4(x',x,y_1,y_2)}}\nonumber \\
& &\ \ \ \ +\cdots\ .
\eea
\par
Calculating the derivative $\bar\delta/\bar\delta x^-$, bringing
the result to the left and completing in the first, second and third
terms of the right-hand side of Eq. (\rf{ae7}) the functions
$S$, $S_{(2)}$ and $S_{(3)}$, respectively, we obtain:
\bea \lb{ae8}
& &\frac{\bar\delta S(x,x')}{\bar\delta x^{\mu -}}=
\frac{\bar\delta F_{2}(x',x)}{\bar\delta x^{\mu -}}\,S(x,x')
+S_0(x,y_1)\gamma^{\alpha_1}\bigg[\Big(
\frac{\bar\delta F_3(x',x,y_1)}{\bar\delta x^{\mu -}}-
\frac{\bar\delta F_2(x',x)}{\bar\delta x^{\mu -}}\Big)
\frac{\bar\delta}{\bar\delta y_1^{\alpha_1 -}}\nonumber \\
& &\ \ \ \ +\frac{\bar\delta^2 F_3(x',x,y_1)}{\bar\delta x^{\mu -}
\bar\delta y_1^{\alpha_1 -}}\bigg]S_{(2)}(y_1,x';x)
\nonumber \\
& &\ \ \ \ +S_0(x,y_1)\gamma^{\alpha_1}S_0(y_1,y_2)\gamma^{\alpha_2}
\bigg[\Big(\frac{\bar\delta F_4(x',x,y_1,y_2)}
{\bar\delta x^{\mu -}}-\frac{\bar\delta F_3(x',x,y_1)}
{\bar\delta x^{\mu -}}\Big)\frac{\bar\delta}{\bar\delta y_1^{\alpha_1 -}}
\frac{\bar\delta}{\bar\delta y_1^{\alpha_2 -}}\nonumber \\
& &\ \ \ \ +\Big(\frac{\bar\delta^2 F_4}{\bar\delta x^{\mu -}
\bar\delta y_1^{\alpha_1 -}}-\frac{\bar\delta^2 F_3}{\bar\delta x^{\mu -}
\bar\delta y_1^{\alpha_1 -}}\Big)
\frac{\bar\delta}{\bar\delta y_2^{\alpha_2 -}}\nonumber \\
& &\ \ \ \ +\frac{\bar\delta^2 F_4}{\bar\delta x^{\mu -}
\bar\delta y_2^{\alpha_2 -}}\frac{\bar\delta}
{\bar\delta y_1^{\alpha_1 -}}
+\frac{\bar\delta^3 F_4}{\bar\delta x^{\mu -}
\bar\delta y_1^{\alpha_1 -}\bar\delta y_2^{\alpha_2 -}}\bigg]
S_{(3)}(y_2,x';x,y_1).
\eea
(Higher-order terms in the derivatives are neglected.)
The operators $\bar\delta/\bar\delta y_1^{\alpha_1 -}$ and
$\bar\delta/\bar\delta y_2^{\alpha_2 -}$ acting on $S_{(2)}$ and 
$S_{(3)}$, respectively, can be replaced in terms of the 
corresponding equation of motion operators and delta functions and
then integrations by parts are carried out; simailarly, the
operator $\bar\delta/\bar\delta y_1^{\alpha_1 -}$ acting on
$S_{(3)}$ can be replaced by 
$\partial/\partial y_1^{}-\bar\delta/\bar\delta y_1^+$ followed by
an integration by parts. One finds at the end:
\bea \lb{ae9}
& &\frac{\bar\delta S(x,x')}{\bar\delta x^{\mu -}}=
\frac{\bar\delta F_2(x',x)}{\bar\delta x^{\mu -}}\,S(x,x')
-\frac{\bar\delta^2 F_3(x',x,y_1)}{\bar\delta x^{\mu -}
\bar\delta y_1^{\alpha_1 +}}\,S_0(x,y_1)\,\gamma^{\alpha_1}\,
S_{(2)}(y_1,x';x)
\nonumber \\
& &\ \ \ \ +\frac{\bar\delta^3 F_4(x',x,y_1,y_2)}{\bar\delta x^{\mu -}
\bar\delta y_1^{\alpha_1 +}\bar\delta y_2^{\alpha_2 +}}\,S_0(x,y_1)\,
\gamma^{\alpha_1}\,S_0(y_1,y_2)\,\gamma^{\alpha_2}\,S_{(3)}(y_2,x';x,y_1)
\nonumber\\
& &\ \ \ \ +\frac{\bar\delta^2 F_4}{\bar\delta x^{\mu -}
\bar\delta y_2^{\alpha_2 +}}\,
\frac{\bar\delta F_4}{\bar\delta y_1^{\alpha_1 +}}
S_0(x,y_1)\,\gamma^{\alpha_1}\,S_0(y_1,y_2)\,\gamma^{\alpha_2}\,
S_{(3)}(y_2,x';x,y_1).
\eea
This result could also have been obtained by using the method of
Secs. \rf{s4} and \rf{s5}, but using for the expansion of $S(A)$ 
Eq. (\rf{3e5}), instead of (\rf{3e11}). It is sufficient for this 
to replace in Eq. (\rf{5e7}) $\bar\delta S/\bar\delta y_i^{}$ by 
zero and $S$ in internal lines by $S_0$.
\par
If we had started for the previous calculations from representations
(\rf{ae1}) and (\rf{ae2}), we would have found in a first stage the 
terms of Eq. (\rf{ae9}) with a remainder containing difference terms 
of the type $(\bar\delta^2 F_4/\bar\delta x^-\bar\delta y_1^+
-\bar\delta^2 F_3/\bar\delta x^-\bar\delta y_1^+)
\bar\delta/\bar\delta y_2^+e^{F_4}$, 
$(\bar\delta F_4/\bar\delta x^--\bar\delta F_3/\bar\delta x^-)
\bar\delta^2/\bar\delta y_1^+\bar\delta y_2^+e^{F_4}$, etc. Integrations
by parts show that the remainder is null. For this, one must proceed in 
two steps. First one converts $\bar\delta/\bar\delta y_2^+$ into
$\bar\delta/\bar\delta y_2^-$ by the formula  
$\bar\delta/\bar\delta y_2^+=\partial/\partial y_2^{}-
\bar\delta/\bar\delta y_2^-$ and an integration by parts is done with 
respect to the total derivative of $y_2$. Second, for the term containing 
$\bar\delta/\bar\delta y_2^-$, one completes the exponential function 
with the multiplicative propagator $S_0(y_2,x')$
into $S_{(3)}$ and then the operator $\bar\delta/\bar\delta y_2^-$ is
replaced in terms of the equation of motion operator and a delta function
and a new integration by parts is done. The net result is zero. This
method of calculation is repeated for all parts of the remainder. 
One thus finds the same result (\rf{ae9}) from both representations 
(\rf{ae1})-(\rf{ae2}) and (\rf{ae3})-(\rf{ae4}). 
\par
Inspection of Eq. (\rf{ae9}) shows that the last term is of the
nested type, with $\bar\delta F_4/\bar\delta y_1^+$ representing
a kind of self-energy insertion on the line $xy_2^{}$. It should
naturally be grouped with the free propagator $S_0$ appearing in
front of $S_{(2)}$ (with a relabelling of the variables $y_1^{}$
and $y_2^{}$) to produce the full Green's function $S$ 
[Eq. (\rf{ae1})] (at the present level of approximation). 
Nevertheless, we are faced with the phenomenon of the Wilson loop 
background effect: the various factors that appear in the last term 
of Eq. (\rf{ae9}) involve $F_4$ and not $F_3$ which is the required 
function in the next-to-leading term of $S$. The recombinations that 
we can do to reconstruct full Green's functions in internal lines
leave at the end remainders of the nested type.
\par
Before proceeding to a recombination of the above factors in the
general case, let us consider first the particular case of
two-dimensional QCD in the quenched approximation (quark loops
neglected) \cite{kkk}. We assume that the Wilson loop contours 
that mainly contribute to the internal integrations are simple
and convex, in particular without self-intersections. In that case,
the logarithm of the Wilson loop average is given by the area of 
the surface delimited by the closed contour and the functions $F_i$ are
proportional to such areas. Furthermore, it is evident that these
areas are separable into smaller ones. For the specific case above,
we have $F_4(x',x,y_1,y_2)=F_3(x',x,y_2)+F_3(x,y_1,y_2)$, which also
implies $\bar\delta F_4(x',x,y_1,y_2)/\bar\delta y_1^+=
\bar\delta F_3(x,y_1,y_2)/\bar\delta y_1^+$,
$\bar\delta^2 F_4(x',x,y_1,y_2)/\bar\delta x^-\bar\delta y_2^+
=\bar\delta^2 F_3(x',x,y_2)/\bar\delta x^-\bar\delta y_2^+$, etc.     
With such decompositions, one easily transforms the last term of
Eq. (\rf{ae9}) into a form that is absorbed by the term containing
$S_{(2)}$ to yield in front of it the full Green's function $S$. Replacing 
in the remaining term containing $S_{(3)}$ the free propagators $S_0$
by $S$ (valid at the present level of approximation), one finds
\bea \lb{ae10}
& &\frac{\bar\delta S(x,x')}{\bar\delta x^{\mu -}}=
\frac{\bar\delta F_2(x',x)}{\bar\delta x^{\mu -}}\,S(x,x')
-\frac{\bar\delta^2 F_3(x',x,y_1)}{\bar\delta x^{\mu -}
\bar\delta y_1^{\alpha_1 +}}\,S(x,y_1)\,\gamma^{\alpha_1}\,
S_{(2)}(y_1,x';x)
\nonumber \\
& &\ \ \ \ +\frac{\bar\delta^3 F_4(x',x,y_1,y_2)}{\bar\delta x^{\mu -}
\bar\delta y_1^{\alpha_1 +}\bar\delta y_2^{\alpha_2 +}}\,S(x,y_1)\,
\gamma^{\alpha_1}\,S(y_1,y_2)\,\gamma^{\alpha_2}\,S_{(3)}(y_2,x';x,y_1),
\eea
which is an expansion with full Green's functions in internal lines
and kernels of the irreducible type without nested diagrams.
\par
In four dimensions, the above decompositions of the Wilson loop
averages are not generally valid and one has to evaluate the
remainder with respect to Eq. (\rf{ae10}). To this end, we
complete in Eq. (\rf{ae9}) the factor $S_0(x,y_1)$ in the term 
containing $S_{(2)}$ into $S(x,y_1)$ and isolate the rest, which is
equal to
\be \lb{ae11}
-S_0(x,z_1)\gamma^{\beta_1}S_0(z_1,y_1)
\frac{\bar\delta F_3(y_1,x,z_1)}{\bar\delta z_1^{\beta_1 +}}
e^{{\displaystyle F_3(y_1,x,z_1)}}\gamma^{\alpha_1}
\frac{\bar\delta^2 F_3(x',x,y_1)}{\bar\delta x^{\mu -}
\bar\delta y_1^{\alpha_1 +}}S_{(2)}(y_1,x';x)
\ee
and which we write in the form
\bea \lb{ae12}
& &-S_0(x,z_1)\gamma^{\beta_1}S_0(z_1,y_1)
\frac{\bar\delta F_3(y_1,x,z_1)}{\bar\delta z_1^{\beta_1 +}}
e^{{\displaystyle F_3(y_1,x,z_1)}}\gamma^{\alpha_2}
\delta^4(y_1-y_2)\nonumber \\
& &\ \ \ \ \ \ \ \ \times
\frac{\bar\delta^2 F_4(x',x,y_1,y_2)}{\bar\delta x^{\mu -}
\bar\delta y_2^{\alpha_2 +}}S_{(3)}(y_2,x';x,y_1).
\eea
Writing the delta-function in the form 
$i\delta^4(y_1-y_2)=(i\gamma.\partial_{(y_1)}-m)S_0(y_1,y_2)$ and 
making an integration by parts with respect to $y_1$ and neglecting 
higher-order derivative terms, we obtain
\be \lb{ae13}
-S_0(x,y_1)\gamma^{\alpha_1}S_0(y_1,y_2)
\frac{\bar\delta F_2(y_1,x)}{\bar\delta y_1^{\alpha_1 +}}
e^{{\displaystyle F_2(y_1,x)}}\gamma^{\alpha_2}
\frac{\bar\delta^2 F_4(x',x,y_1,y_2)}{\bar\delta x^{\mu -}
\bar\delta y_2^{\alpha_2 +}}S_{(3)}(y_2,x';x,y_1).
\ee
In the first terms we recognize the dominant pieces 
of the product $\frac{\bar\delta S(x,y_1)}{\bar\delta y_1^{\alpha_1 +}}
\gamma^{\alpha_1}\,S(y_1,y_2)$. Using the equation of motion of $S$
and making again an integration by parts with respect to $y_1^{}$
(neglecting higher-order derivatives) we find the final expression
\be \lb{ae14}
S(x,y_1)\gamma^{\alpha_1}
\frac{\bar\delta S(y_1,y_2)}{\bar\delta y_1^{\alpha_1 -}}
\gamma^{\alpha_2}
\frac{\bar\delta^2 F_4(x',x,y_1,y_2)}{\bar\delta x^{\mu -}
\bar\delta y_2^{\alpha_2 +}}S_{(3)}(y_2,x';x,y_1),
\ee
which, when grouped with the last term of Eq. (\rf{ae9}), in which
the $S_0$s may be replaced by $S$s, yields the nested piece of
Eq. (\rf{5e7}) accompanying $S_{(3)}$:
\be \lb{ae15}
\frac{\bar\delta^2 F_4(x',x,y_1,y_2)}{\bar\delta x^{\mu -}
\bar\delta y_2^{\alpha_2 +}}S(x,y_1)\gamma^{\alpha_1}
\Big(\frac{\bar\delta S(y_1,y_2)}{\bar\delta y_1^{\alpha_1 -}}
+S(y_1,y_2)\frac{\bar\delta F_4}{\bar\delta y_1^{\alpha_1 +}}\Big)
\gamma^{\alpha_2}S_{(3)}(y_2,x';x,y_1).
\ee
\par
We thus recover, together with the terms of Eq. (\rf{ae10}), the first
terms of Eq. (\rf{5e7}), up to $S_{(3)}$. The calculation could be
continued to higher-orders in the derivative terms, but it rapidly
becomes complicated and loses interest for practical applicability.
The method is useful for low-order perturbative calculations 
and for analysis of general qualitative properties.
\par

\section{Quark self-energy} \lb{ap2}
\setcounter{equation}{0}

In order to construct the quark self-energy function directly,
without having recourse to the explicit expression of the
Green's function $S$, we start from its relationship with the 
derivative terms of $S$. The self-energy function, which we designate
by $\Sigma$, is defined from the inverse of the Green's function:
\be \lb{be1}
iS^{-1}(x,x')=(i\gamma.\partial_{(x)}-m)\,\delta^4(x-x')
-\Sigma(x,x').
\ee
Comparison with the equations of motion (\rf{3e8a}) and (\rf{3e8b})
yields:
\be \lb{be2}
\Sigma(x,x'')\,S(x'',x')=
i\gamma^{\mu}\,\frac{\bar\delta S(x,x')}{\bar\delta x^{\mu -}},
\ \ \ \ \ S(x,x'')\,\Sigma(x'',x')=
-i\frac{\bar\delta S(x,x')}{\bar\delta x^{'\nu +}}\,\gamma^{\nu}.
\ee
\par
Letting $S^{-1}$ act on these equations, one ends up with the equation 
for $\Sigma$:
\be \lb{be3}
\Sigma(x,x')=-i\gamma^{\mu}\,\frac{\bar\delta^2 S(x,x')}
{\bar\delta x^{\mu -}\bar\delta x^{'\nu +}}\,\gamma^{\nu}
+i\Sigma(x,y)\,S(y,y')\,\Sigma(y',x').
\ee
The second-order derivative 
$\bar\delta^2 S(x,x')/\bar\delta x^{\mu -}\bar\delta x^{'\nu +}$
corresponds to a generalization of the first-order derivatives
$\bar\delta S/\bar\delta x^{\pm}$, defined in Eqs. (\rf{2e18}) and 
(\rf{2e5a})-(\rf{2e5b}), where now the two derivations act on
the same segment $xx'$; in this case, one has also to take into
account the contributions coming from coincident points. 
Explicitly, one has:
\bea \lb{be4}
& &\frac{\bar\delta^2 U(x',x)}
{\bar\delta x^{\mu -}\bar\delta x^{'\nu +}}=
(ig)^2\,\int_0^1d\lambda d\lambda'(1-\lambda)\lambda'\,
\Big[U(1,\lambda')z^{'\beta}(\lambda')F_{\beta\nu}(z(\lambda'))
U(\lambda',\lambda)\nonumber \\
& &\ \ \times z^{'\alpha}(\lambda)F_{\alpha\mu}(z(\lambda))
U(\lambda,0)+U(1,\lambda)z^{'\alpha}(\lambda)F_{\alpha\mu}(z(\lambda))
U(\lambda,\lambda')z^{'\beta}(\lambda')F_{\beta\nu}(z(\lambda'))
U(\lambda',0)\Big]\nonumber \\
& &\ \ +ig\int_0^1d\lambda(1-\lambda)\lambda\,U(1,\lambda)
z^{'\alpha}(\lambda)\Big(\nabla_{\nu}F_{\alpha\mu}(z(\lambda))\Big)
U(\lambda,0)\nonumber \\
& &\ \ +ig\int_0^1d\lambda(1-\lambda)\,U(1,\lambda)
F_{\nu\mu}(z(\lambda))U(\lambda,0),
\eea  
where $z'(\lambda)=\frac{\partial z}{\partial \lambda}=x'-x$ and
$\nabla$ is the covariant derivative, $(\nabla F)=(\partial F)+ig[A,F]$.
The calculation was done by first deriving with respect to $x$ and then
with respect to $x'$. Had we interchanged the orders of 
derivation, the last two terms would be modified in the
following way: 
$z^{'\alpha}\nabla_{\nu}F_{\alpha\mu}\rightarrow 
z^{'\alpha}\nabla_{\mu}F_{\alpha\nu}$, 
$(1-\lambda)F_{\nu\mu}\rightarrow -\lambda F_{\mu\nu}$. 
The two expressions are, however, equivalent, since the orders
of derivation are irrelevant, due to the Bianchi identity satisfied
by $F$. This can be checked directly by taking the difference of
the two expressions.
\par
The two derivative term (\rf{be4}) requires a careful
treatment, since it contains divergences or singularities not 
present in one derivative terms. This is the case for the 
trace of the tensor (\rf{be4}); for neighboring points in the
expression inside the brackets, the two $F$s lead to a 
divergence, even when short-distance perturbative interactions are 
ignored \cite{lsw,js}; still for the trace, the term with the 
covariant derivative reduces to the gluon equation of motion operator
and hence yields a delta-function (plus a quark current); these 
singular terms should be grouped together to set up a regularized 
form of the corresponding quantities. Concerning the traceless 
part of the coincident points contribution (the last two terms of
Eq. (\rf{be4})), we observe that it is of order $g$ and not $g^2$;
this implies that when expanding equation (\rf{be3}) in terms of
derivatives of Wilson loop averages, one should count the latter 
contribution as a one derivative term. Finally, the role of the last 
term of Eq. (\rf{be3}) is to cancel similar reducible type terms 
that might emerge from the expansion of the second-order derivative
piece. Nevertheless, because of existing background effects, as in
the case of nested diagrams met in Sec. \rf{s5}, the cancellations
are only partial.
\par
The above features make the direct treatment of the self-energy 
function rather intricate and less appealing than that of the Green's 
function itself. This underlines the fact that proper vertices 
do not seem to play a primary role in the present approach.
We shall not pursue any longer here the study of the self-energy 
function.
\par

\end{document}